\documentclass[preprint,3p,12pt,authoryear]{elsarticle}

\usepackage{amssymb}
\usepackage{amsmath}
\usepackage{bm}
\usepackage{booktabs}
\usepackage{caption}
\usepackage{graphicx}
\usepackage{multirow}
\usepackage{tabularx}
\usepackage{color}

\journal{Expert Systems with Applications}

\begin{document}

\begin{frontmatter}



\title{H2-MARL: Multi-Agent Reinforcement Learning for Pareto Optimality in Hospital Capacity Strain and Human Mobility during Epidemic} 


\author[1]{Xueting Luo} 
\ead{2333097@tongji.edu.cn}
\author[1,7]{Hao Deng\corref{cor1}}
\ead{denghao1984@tongji.edu.cn}
\author[2]{Jihong Yang}
\ead{yangjihong@mail.clspi.org.cn}
\author[3]{Yao Shen}
\ead{eshenyao@tongji.edu.cn}
\author[4]{Huanhuan Guo}
\ead{guohuan2009@126.com}
\author[1]{Zhiyuan Sun}
\ead{2433274@tongji.edu.cn}
\author[5]{Mingqing Liu}
\ead{ml2176@cam.ac.uk}
\author[6]{Jiming Wei}
\ead{mjway@qq.com}
\author[1,7]{Shengjie Zhao}
\ead{shengjiezhao@tongji.edu.cn}

\affiliation[1]{organization={School of Computer Science and Technology, Tongji University},
            city={Shanghai},
            postcode={201804}, 
            country={China}}

\affiliation[2]{organization={China Land Surveying and Planning Institute},
            city={Beijing},
            postcode={100035}, 
            country={China}}

\affiliation[3]{organization={College of Architecture and Urban Planning, Tongji University},
            city={Shanghai},
            postcode={200092}, 
            country={China}}

\affiliation[4]{organization={Chongqing Institute of Planning and Natural Resources Investigation and Monitoring},
            city={Chongqing},
            postcode={401121}, 
            country={China}}

\affiliation[5]{organization={Li-Fi Research and Development Centre, University of Cambridge},
            city={Cambridge},
            postcode={CB3 0FA}, 
            country={UK}}

\affiliation[6]{organization={Guangdong Urban-rural Planning and Design Research Institute Technology Group Co., Ltd.},
            city={Guangzhou},
            postcode={510290}, 
            country={China}}

\affiliation[7]{organization={Engineering Research Center of Key Software Technologies for Smart City Perception and Planning, Ministry of Education},
            city={Shanghai},
            postcode={201804}, 
            country={China}}

\cortext[cor1]{Corresponding author}

\begin{abstract}
The necessity of achieving an effective balance between minimizing the losses associated with restricting human mobility and ensuring hospital capacity has gained significant attention in the aftermath of COVID-19. Reinforcement learning (RL)-based strategies for human mobility management have recently advanced in addressing the dynamic evolution of cities and epidemics; however, they still face challenges in achieving coordinated control at the township level and adapting to cities of varying scales. To address the above issues, we propose a multi-agent RL approach that achieves Pareto optimality in managing hospital capacity and human mobility (H2-MARL), applicable across cities of different scales. We first develop a township-level infection model with online-updatable parameters to simulate disease transmission and construct a city-wide dynamic spatiotemporal epidemic simulator. On this basis, H2-MARL is designed to treat each division as an agent, with a trade-off dual-objective reward function formulated and an experience replay buffer enriched with expert knowledge built. To evaluate the effectiveness of the model, we construct a township-level human mobility dataset containing over one billion records from four representative cities of varying scales. Extensive experiments demonstrate that H2-MARL has the optimal dual-objective trade-off capability, which can minimize hospital capacity strain while minimizing human mobility restriction loss. Meanwhile, the applicability of the proposed model to epidemic control in cities of varying scales is verified, which showcases its feasibility and versatility in practical applications.
\end{abstract}





\end{frontmatter}


\section{Introduction}
Are we ready for the next epidemic? The experience of COVID-19 has shown that frequent inter-regional mobility under urbanization can accelerate the spread of epidemics, posing significant threats to public health and economic activities. Human mobility restrictions, such as travel restrictions, facility closures, and traffic controls \citep{fenichel2013economic,zhou2020effects,xiang2023impact}, are considered effective measures to slow the transmission of epidemics by reducing the frequency of human contact and curbing the spread of infection \citep{liu2020covid}. However, developing effective strategies is challenging due to the dynamic nature of disease transmission and the need to balance multiple conflict factors, including hospital capacity strain \citep{atalan2020lockdown, lin2022hospital} and mobility restriction loss \citep{haw2022optimizing}. In this context, precise epidemic simulation and effective mobility restriction strategies become particularly crucial.

Existing epidemic simulation models can be categorized into mathematical models, complex network models, and agent-based models \citep{duan2015mathematical}. Mathematical models \citep{eseghir2013branching, xiao2011progress} simplify the epidemic transmission process, which limits their ability to capture the complexity of disease spread. Complex network models \citep{liu2024review} effectively represent heterogeneous population structures and interaction patterns. However, they face challenges in capturing the spatiotemporal dynamics of epidemic transmission. Agent-based models \citep{silva2020covid} utilize complex networks to represent agent interactions while employing mathematical methods to guide behavior. Nonetheless, they struggle with substantial computational resources and large, high-quality datasets. Furthermore, existing models have limitations in parameter settings. The method of presetting parameters \citep{meltzer2016modeling, merler2015spatiotemporal, chowell2016characterizing} may introduce biases in epidemic simulations and evaluations of control measures. While some studies \citep{lai2020early, zhao2020modelling, chong2020novel} attempt to update parameters by utilizing predefined time functions or collecting complete transmission chains, they still encounter challenges in obtaining actual data and addressing the time-lag issue. Thus, an effective epidemic model that simulates dynamic changes on spatiotemporal scales is required, which will support the restriction strategies.

In the research of human mobility management strategies during epidemics, traditional model predictive control (MPC) models \citep{kohler2018dynamic, she2022learning, liu2021use} depend on precise mathematical models and predefined parameters for prediction and optimization. However, they exhibit limitations in addressing the highly dynamic and complex nonlinear nature of epidemic spread. Reinforcement Learning (RL) \citep{rao2024modeling, song2020reinforced, feng2023contact}, as an adaptive intelligent optimization method, obtains the optimal strategies based on environmental feedback, particularly demonstrating significant advantages in dynamic multi-objective epidemic control. Nonetheless, RL in mobility restriction still faces challenges, in achieving fine-grained collaborative restrictions at the township-level administrative division (AD) and addressing the adaptability to cities with varying scales. These limitations promote researchers to explore Multi-Agent Reinforcement Learning (MARL) as a more robust solution. MARL employs multiple regional agents to collaborate, with each agent generating differentiated strategies based on local observation \citep{selvi2021deep, zhang2022neighborhood}. However, challenges remain in terms of information sharing, strategy coordination, and learning efficiency.

To address the above challenges, we introduce an epidemic spatiotemporal simulator that can update parameters online, where a dynamic susceptible-infected-hospitalized-removed (D-SIHR) is employed for infection simulation. Besides, considering the issue of dual-objective trade-off and exploration effectiveness, we develop a \textit{\textbf{M}ulti-\textbf{A}gent \textbf{R}einforcement \textbf{L}earning for Pareto Optimality in \textbf{H}ospital Capacity Strain and \textbf{H}uman Mobility} (H2-MARL) model. Moreover, since there is no off-the-shelf real-world human mobility dataset at the township level covering cities of varying scales for mobility restriction research, we establish a mobility dataset using mobile phone signaling data. This paper presents the following contributions:

\begin{itemize}
    \item {We propose a data-driven spatiotemporal environment simulator utilizing the D-SIHR infection model to capture the infection dynamics at the township level, which provides precise simulations of epidemic spread and control across diverse urban settings for more targeted and fine-grained analysis. A time-lag-free online parameter updating method is designed and incorporated to adjust infection rates, ensuring more accurate reflections of epidemic conditions.}
    \item {We develop a mobility restriction strategy model H2-MARL to achieve Pareto optimality between hospital capacity strain and human mobility, where a novel reward function is designed to dynamically adjust weights towards two objectives trade-off employing the entropy weight method. By integrating expert experience replay buffers and heuristic spatial pruning, H2-MARL improves the exploration process, preventing inefficient or suboptimal strategies.}
    \item {We establish a large-scale origin-destination (OD) dataset comprising over 1 billion historical records from four typical cities of varying scales to address the lack of publicly available mobility datasets for township-level ADs. This dataset supports more refined analyses of mobility constraints and serves as a crucial foundation for developing and evaluating mobility restriction strategies.}
    \item {We conduct comprehensive experiments to validate the proposed D-SIHR infection model and the H2-MARL strategy. Through comparisons with existing models, we assess effectiveness, scalability, and adaptability, demonstrating the superior performance of H2-MARL in online dual-objective decision-making, and confirming the potential utility in epidemic control and epidemic prevention zone planning.}
\end{itemize}

The remainder of the manuscript is as follows. Section \ref{section:related_work} provides related work on epidemic compartmental model and RL-based methods in epidemic human mobility restriction. In Section \ref{section:methods}, we present a detailed description of the model structure and algorithm flow. Section \ref{section:experiments} mainly discusses the datasets, evaluation criteria, and experimental outcomes, followed by comparative analysis. Finally, in Section \ref{section:conclision}, we give conclusions and discuss future extensions.

\section{Related work}
\label{section:related_work}
In this section, we will review some representative work in the field of epidemic compartmental models and RL-based methods in epidemic human mobility restriction.

\subsection{Epidemic compartmental model}
The epidemic compartmental models, as primary mathematical tools for studying and simulating disease spread, have been widely applied and researched in epidemiology and public health. One of the most classical compartmental models is the susceptible-infected-removed (SIR) model \citep{kermack1927contribution}, which categorizes the population into homogeneous susceptible ($S$), infected ($I$), and recovered ($R$) individuals, describing the transmission process of epidemics within the population. The SIR model is expressed as a set of ordinary differential equations:
\begin{equation}
    \begin{aligned}
        \frac{dS}{dt} &= -\beta \frac{S I}{N}, \\
        \frac{dI}{dt} &= \beta \frac{S I}{N} - \gamma I, \\
        \frac{dR}{dt} &= \gamma I,
    \end{aligned}
    \label{eq: SIR}
\end{equation}
where $S$, $I$, and $R$ represent the number of susceptible, infected, and recovered individuals, respectively. $N$ denotes the total population size, where $N = S + I + R$. $\beta$ indicates the infection rate, and $\gamma$ represents the recovery rate. The SIR model ignores the dynamics of transmission \citep{cao2020COVIDcomplex}, including human mobility, policy restrictions, and viral mutations.

As research progresses, epidemic compartment models have made many extensions from different aspects. For instance, the SIR model has been enhanced by incorporating factors such as latent period \citep{liu2020covid}, vaccination \citep{dutta2024periodic}, age \citep{davies2020age}, and geographic information \citep{boscheri2021modeling}. In particular, to measure the hospital capacity resources, Pan et al. in \citep{pan2021covid} proposed a susceptible-infected-hospitalized-recovered (SIHR) model, which introduces a new hospitalization state, denoted by $H$, between the $I$ and $R$ state in the SIR model. It is easy to see that compartmental models are highly sensitive to parameter setting, especially the infection rate $\beta$ in (\ref{eq: SIR}), which directly influences the result of the model. However, traditional models \citep{bacaer2020pic, hespanha2021forecasting} often overlook this aspect, which is typically characterized by constant parameters and select the optimal values by model fitting. This approach may fail to capture the evolution of the disease dynamics due to changes in social behavior, non-pharmaceutical interventions, and testing rates. Some studies \citep{calafiore2020time, godio2020seir, al2021sir} have expressed the time-varying parameters through a linear combination of pre-specified time functions, with the coefficients of these linear combinations identified from data. Kiamari et al. in \citep{kiamari2020covid} took a simpler approximation method, which sets the infection rate as a time function of the infectors. The model in \citep{ihme2021modeling} considered a time-varying infection rate as a linear combination of a set of explanatory covariates that include seasonality, mobility, and mask use. Time function permits more realistic smooth variations of the parameters, but makes the outcomes highly dependent on the choice of the basis functions, which must be pre-specified and not learned from data.

Learning parameters from real-time infection data is considered more accurate and reasonable \citep{jung2020real}. It requires counting the number of individuals directly infected by each infector in the next generation to update the infection rate \citep{ward2024real, da2021covid, arroyo2021tracking}. However, it is challenging to obtain the complete transmission chain of the epidemic in practice. Wallinga et al. in \citep{wallinga2004different} proposed a likelihood-based estimation method that requires obtaining infection data over a complete serial interval time period following the current time, resulting in a lag in the estimates. Cauchemez et al. in \citep{cauchemez2006real} later revised this method to address the time-lag issue. Thompson et al. in \citep{thompson2019improved} integrated data from known primary and secondary case pairs to directly estimate the serial interval, addressing the uncertainty in the early stages of an outbreak. These methods \citep{zhao2020modelling, chong2020novel, huang2020dynamic} rely on observations of transmission pairs and time series data to estimate the serial interval, followed by calculating infection rates. However, it is challenging to fully capture actual transmission pairs, necessitating modeling of the infection process to fit the serial interval time. In our work, we estimate the infection rate by evaluating the transmission capability of new infectors, and correct for potential underestimation due to time lag by the serial interval distribution.

\subsection{RL-based methods in epidemic human mobility restriction}
RL is regarded as an effective approach for addressing the challenges of human mobility restrictions during epidemic outbreaks \citep{qin2024applying, bampa2022epidrlearn}. It can generate dynamic and adaptive strategies to mitigate the spread of diseases based on environmental information. Epidemic control typically involves multiple conflicting objectives \citep{balcells2024multi, ahmed2020epidemic}, such as reducing infections, minimizing the socio-economic impact of mobility restrictions, and alleviating psychological burden. To address this challenge, some studies have proposed RL-based multi-objective optimization methods. Tang et al. in \citep{tang2022bi} combined the two conflicting objectives to form a single-objective optimization problem based on the weight-sum method. Wan et al. in \citep{wan2021multi} proposed a model-based planning algorithm to learn the optimal restriction strategy, for each given weight between the two competing objectives. The above methods of setting weights still have research gaps. They overlooked the relative importance of objectives in the epidemic scenario. Additionally, the weights need to be adjusted adaptively to meet the preferences of the environment.

Furthermore, researchers have explored the potential of MARL algorithms for optimizing epidemic mobility restrictions \citep{shaik2024adaptivemultiagentdeepreinforcement, hurtado2023quarantine}. MARL has better regional collaboration capability, and its strategies offer enhanced scalability for regional agents \citep{carley2006biowar, valianti2024cooperative}. Hurtado et al. in \citep{hurtadoand} implemented an offline MARL algorithm to achieve individual collaborative decision-making, controlling access to urban POIs to reduce exposure risk. Zhang et al. in \citep{zhang2022neighborhood} proposed fully cooperative agents to coordinate regional lockdown policies, enabling dynamic responses to surges in infections while minimizing broader societal impacts. For scenarios with limited urban resources, Seid et al. in \citep{seid2023multiagent} introduced semi-cooperative, semi-competitive agents, where different regions compete for scarce resources while also balancing the overall cost of epidemic control.

In summary, the challenges of the existing epidemic mobility restriction models lie in the lack of adaptive weight adjustment methods for balancing conflicting objectives. Considering the advantages of MARL in inter-regional collaboration, we propose a MARL-based model for Pareto optimality in hospital capacity strain and human mobility, achieving the optimal restriction strategy for multi-AD cooperation at the township level.

\section{Methods}

\begin{figure}
	\centering
	\includegraphics[width=0.8\linewidth]{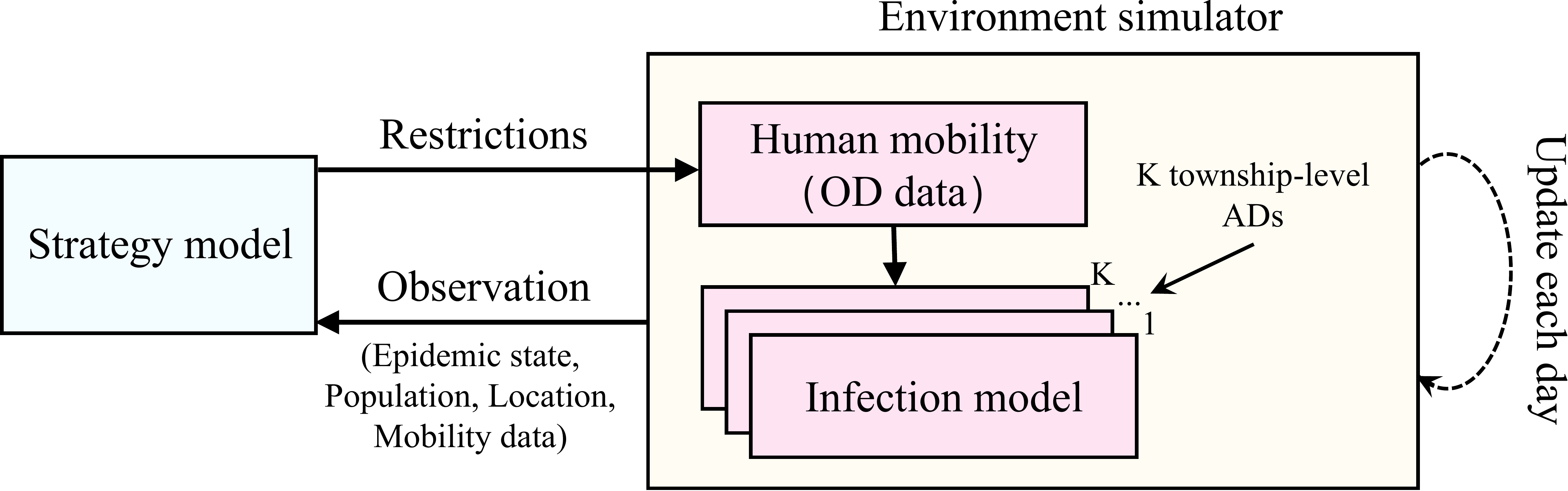}
	\caption{Block diagram of the proposed environment simulator and interaction with mobility restriction strategy.}
	\label{fig:simulator_framework}
\end{figure}

\label{section:methods}
In this section, we present the proposed H2-MARL model in two parts. First, we develop the D-SIHR based environment simulator that precisely models the epidemic transmission process under human mobility across multiple ADs in an urban setting. Second, we introduce the design of the H2-MARL human mobility restriction strategy model, including the MARL structure, the dual-objective reward function, and the optimization process.

\subsection{D-SIHR based environment simulator}
The environment simulator with a daily time step provides a virtual environment for simulating the epidemic transmission and evaluating the effectiveness of the strategy. As in Fig. \ref{fig:simulator_framework}, $K$ township-level ADs are assumed to comprise the urban setting, where each township-level AD is the smallest unit with varying population densities and medical resources. The established environment simulator involves 2 blocks: the human mobility block and the infection model for each township-level AD. The denotations used are summarized in Table~\ref{tab:symbol}.

\begin{table}[htbp]
  \caption{Notations used in methods.}\label{tab:symbol}
  \centering
  \resizebox{\textwidth}{!}{
  \begin{tabular}{cc}
  \toprule
  \textbf{Symbols} & \textbf{Instructions} \\
  \midrule
  $\mathbf{M}_d^t$ & The ADs' mobility demand matrix on day $t$\\
  $\mathbf{Q}^t$ & The mobility quota matrix on day $t$ for all ADs \\
  $\mathbf{q}_i^t$ & The $i$-th row vector of the $\mathbf{Q}^t$ \\
  $\mathbf{Q}^{pre}$ & The mobility quota matrix defined by expert policy for all ADs \\
  $\mathbf{M}^t$ & The restricted mobility matrix under quota $\mathbf{Q}^t$ on day $t$ \\
  $m^t_{d, ij}$ & Element of $\mathbf{M}_d^t$, number of population moving from AD $i$ to $j$ in demand on day $t$ \\
  $q^t_{ij}$ & Element of $\mathbf{Q}^t$, the mobility quota from AD $i$ to $j$ \\
  $q^{pre}_{ij}$ & Element of $\mathbf{Q}^{pre}$, the mobility quota defined by expert policy from AD $i$ to $j$ \\
  $m^t_{ij}$ & Element of $\mathbf{M}^t$, number of population moving from AD $i$ to $j$ under $q^t_{ij}$ on day $t$ \\
  $K$ & The number of urban ADs \\
  subscript $d$ & The label that is used to represent the vocabulary ``demand'' \\
  subscript $+$ & The label that is used to indicate the inflow from other ADs into the current AD \\
  \bottomrule
  \end{tabular}}
\end{table}

The human mobility block depicts the population movement between ADs inside a city on day $t$, which is modeled by a mobility demand matrix $\mathbf{M}_d^t$ derived from OD data. $\mathbf{M}_d^t$ is a $K \times K$ matrix, where each element $m_{d, ij}^t$ indicates the number of population moving from AD $i$ to $j$ in demand on day $t$. The human mobility restriction strategy, which is determined by a mobility quota matrix $\mathbf{Q}^t$, contributes to the update of $\mathbf{M}_d^t$ as:
\begin{equation}
    \label{eq:Mt_Qt}
    \begin{aligned}
    \mathbf{M}^t &= \mathbf{M}_{d}^t \odot \mathbf{Q}^t, \\
    \mathbf{Q}^t &= [q_{ij}^t]_{1 \leq i, j \leq K}, \quad q_{ij}^t \in [0,1], 
    \end{aligned}
\end{equation}
where $\odot$ represents the Hadamard product, and $\mathbf{M}^t$ denotes the actual human mobility matrix after applying restriction strategies. The detailed Formulation of $\mathbf{Q}^t$, which determines the actual mobility during the spread of the epidemic, can be found in Section \ref{section:H2-MARL_strategy_model}.

For the infection modeling block, we construct a dynamic spatiotemporal epidemic compartmental infection model, D-SIHR, in each AD.

\subsubsection{D-SIHR model}

\begin{figure}
	\centering
	\includegraphics[width=0.7\linewidth]{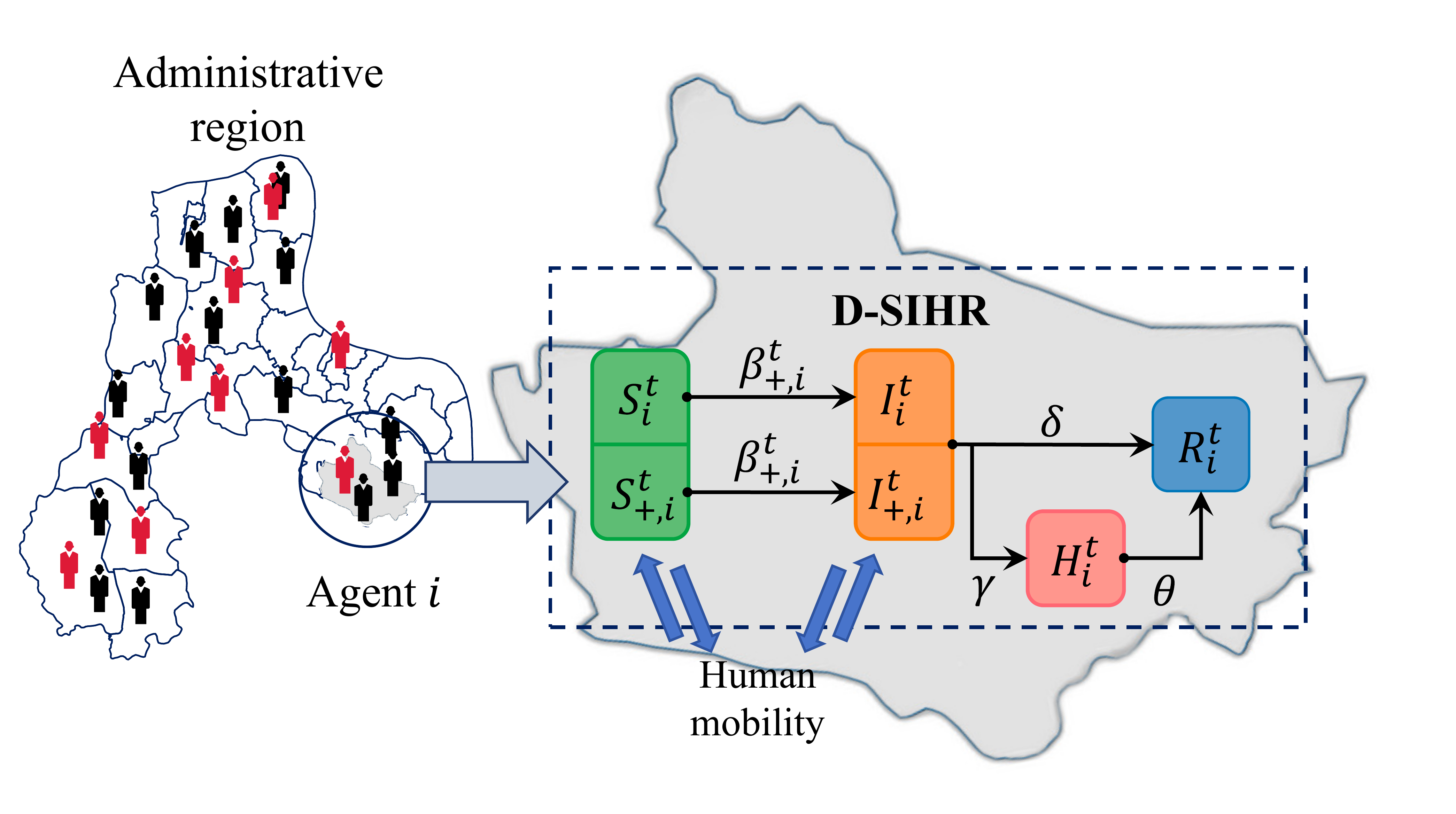}
	\caption{The D-SIHR model for AD $i$ on day $t$.}
	\label{fig:D-SIHR_framework}
\end{figure}

The epidemic transmission under human mobility is depicted in Fig. \ref{fig:D-SIHR_framework}. For AD $i$ on day $t$, we divide the population into four groups, defined as the susceptible ($S_i^t$), infected ($I_i^t$), hospitalized ($H_i^t$) and removed ($R_i^t$) individuals. We assume that hospitalizations are in quarantine and unable to move, and thus do not contribute to the new infections. The $R_i^t$ represents the final stage of the transmission mechanism. Moreover, many epidemics exhibit self-limiting characteristics \citep{zhu2021effect}, meaning that individuals with mild symptoms can recover through their immune system without special treatment or medical intervention. Therefore, we add a direct transition from the infected state $I^t_i$ to the removed state $R^t_i$. The transitions in the proposed D-SIHR model are defined as follows:
\begin{equation}
    \label{eq: D-SIHR}
    \begin{aligned}
        & S_{i, new}^t = -\frac{\beta_{+, i}^t S_{+, i}^t I_{+, i}^t}{N_{+, i}^t} - \frac{\beta_i^t S_i^t I_i^t}{N_i^t}, \\
        & I_{i, new}^t = \frac{\beta_{+, i}^t S_{+, i}^t I_{+, i}^t}{N_{+, i}^t} + \frac{\beta_i^t S_i^t I_i^t}{N_i^t} - (\gamma + \delta) (I_{+, i}^t + I_i^t), \\
        & H_{i, new}^t = \gamma (I_{+, i}^t + I_i^t) - \theta H_i^t, \\
        & R_{i, new}^t = \delta (I_{+, i}^t + I_i^t) + \theta H_i^t,
    \end{aligned}
\end{equation}
where $S_{i, new}^t$, $I_{i, new}^t$, $H_{i, new}^t$ and $R_{i, new}^t$ represent the changes in the numbers of $S_i^t$, $I_i^t$, $H_i^t$ and $R_i^t$ for AD $i$ on day $t$. The hyperparameters $\gamma$, $\theta$, and $\delta$ are the hospitalized rate, the cure rate, and the self-recovery rate, respectively. Additionally, subscript $+$ represents the inflow from other ADs into the current AD. $\beta_{+, i}^t$ and $\beta_i^t$ denote the epidemic's infection rate of the population inflowing and remaining in AD $i$, respectively.

For simplicity in mathematical expressions, we denote $\mathbf{D}^t_i$ as an epidemic state vector, where $\mathbf{D}^t_i = [S^t_i, \; I^t_i, \; H^t_i, \; R^t_i]$. The update of $\mathbf{D}^{t+1}_i$ is expressed as
\begin{equation}
    \mathbf{D}^{t+1}_i = \mathbf{D}^{t}_i + \sum_{j=1}^{K} \frac{m^t_{ji}}{N_j^t}\mathbf{D}^t_j - \sum_{j=1}^{K} \frac{m^t_{ij}}{N^t_i}\mathbf{D}^t_i + \mathbf{D}^t_{i, new},
    \label{eq:D_func}
\end{equation}
where $N^t_i = S^t_i + I^t_i + H^t_i + R^t_i$ denotes the total population of AD $i$ and $\mathbf{D}^t_{i, new} = [S_{i, new}^t, \; I_{i, new}^t, \\ H_{i, new}^t, \; R_{i, new}^t]$ represents the changes in population due to the spread of epidemic.

\subsubsection{Infection rate calculation}

\begin{figure}
	\centering
	\includegraphics[width=0.5\linewidth]{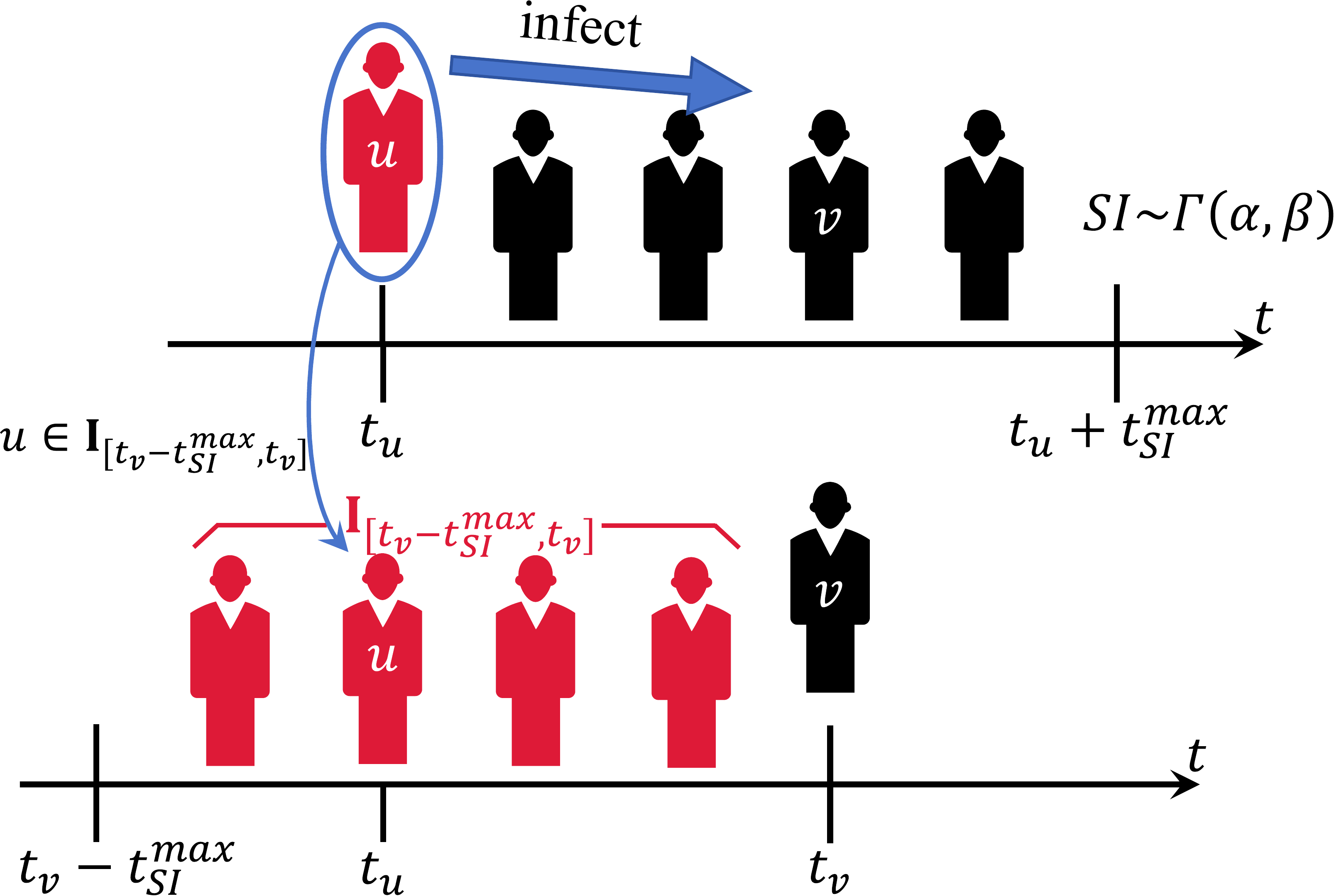}
	\caption{The explanation of $p_{uv}$. Black icons represent healthy individuals, while red icons signify infected individuals. In the top panel, individual $u$ can infect individual $v$ during the maximum serial interval $t_{SI}^{\max}$, with the serial interval following a Gamma distribution $\Gamma(\alpha, \beta)$. In the bottom panel, the set of infectors $\mathbf{I}_{[t_v - t_{SI}^{\max}, t_v]}$ has the ability to infect $v$, $u \in \mathbf{I}_{[t_v - t_{SI}^{\max}, t_v]}$.}
	\label{fig:D-SIHR_Rt}
\end{figure}

The infection rates $\beta_{+, i}^t$ and $\beta_i^t$ in (\ref{eq: D-SIHR}) are time-varying parameters. For simplicity, the infection rates are unified with a denotation $\beta_*^t$, which is obtained through the following equation \citep{gostic2020practical}:
\begin{equation}
    \beta_*^t = \frac{R_{t, *}}{\mathcal{D}},
\end{equation}
where $\mathcal{D}$ represents the effective infectious period \citep{guerra2017basic}, i.e., the number of days during which an infected individual is capable of transmitting the infection. $R_{t, *}$ represents the effective reproduction number on day $t$ \citep{anderson1991infectious},  indicating the average number of individuals infected by each case within the serial interval (SI) \citep{nishiura2020serial}. The serial interval $t_{SI}$ is approximately Gamma distributed, denoted as $t_{SI}\sim\Gamma(\alpha, \beta)$, with $ t_{SI}>0 $. The probability density function is given by $f(t_{SI}, \beta, \alpha)$, where $\alpha$ is the shape parameter and $\beta$ is the scale parameter. We define $t_{{SI}}^{\max}$ as the maximum serial interval time.

$\mathcal{D}$ is a hyperparameter related to the virus properties. Thus, under the restriction measures, the value of $\beta_*^t$ depends on the current $R_{t, *}$, which is defined as
\begin{equation}
    R_{t, *}=\frac{1}{\left| \Delta \mathbf{I}_*^t \right|}\sum_{u\in \Delta \mathbf{I}_*^t}E_u,
\end{equation}
where $\Delta \mathbf{I}_*^t$ denotes the set of new infectors on day $t$, and $E_u$ represents the number of the expected total infections that infector $u$ causes, and $u \in \Delta \mathbf{I}_*^t$. We assume the equal infectious capacity among infectors in an AD, so the effective reproduction number $R_{t, *}$ of the model equals to the expected total number of infections caused by infector $u$ on day $t$, i.e., $R_{t, *} = E_u$ where $t = t_u$, and $t_u$ denotes the day that $u$ is infected.

For infected $u$, $E_u$ is given by
\begin{equation}
    E_u = \sum_{v \in \mathbf{S}_{[t_u, t_u + t_{{SI}}^{\max}]}}p_{uv},
\end{equation}
where $\mathbf{S}_{[t_u, t_u + t_{{SI}}^{\max}]}$ represents the set of all susceptibles during the period $[t_u, t_u + t_{{SI}}^{\max}]$ when $u$ has infectious capacity. For any individual $v \in \mathbf{S}_{[t_u, t_u + t_{{SI}}^{\max}]}$, $p_{uv}$ indicates the probability that $v$ will be infected by $u$, which is proportional to the Gamma function $\Gamma \left(t_v-t_u\right)$.

As shown in Fig. \ref{fig:D-SIHR_Rt}, $p_{uv}$ is obtained by the ratio of the probability that $v$ is infected by $u$ to the total probability of all potential infectors that could infect $v$, which is given by
\begin{equation}
    p_{uv}=\frac{\Gamma \left(t_v-t_u\right)}{\sum_{w \in \mathbf{I}_{[t_v - t^{max}_{SI}, t_v]}}\Gamma\left(t_v-t_w\right)},
\end{equation}
where $\mathbf{I}_{[t_v - t^{max}_{SI}, t_v]}$ denotes the set of all infectors during the period $[t_v - t^{max}_{SI}, t_v]$ that have the ability to infect $v$.

\subsubsection{Model correction}
When the D-SIHR model calculates $R_{t, *}$ on a specified day $T$, the case where $T < t_u + t^{max}_{SI}$ may occur. It implies that potential cases infected by $u$ after $T$ can not be counted, leading to an underestimation of $R_{t, *}$. Therefore, we introduce a corrected value for $R_{t, *}$, denoted as $R_{t, *}^{c}$, which is expressed as
\begin{equation}
    \label{eq: Rt_correction}
    R_{t, *}^{c} = \frac{1}{|\Delta I_*^t|} \sum_{u \in \Delta \mathbf{I}_*^t} (E_u + E_u'),
\end{equation}
\begin{equation}
    E_u' = E_u \times \left(1 - F(T - t_u; \alpha, \beta)\right),
\end{equation}
where $E_u'$ indicates the expected additional number of infectors that might originate from infector $u$ after day $T$. We utilize the cumulative distribution function (CDF) of the $\Gamma(\alpha, \beta)$ distribution, denoted as $F(T - t_u; \alpha, \beta)$, to calculate the cumulative probability of infections occurring within the interval $(t_{u}, T]$. For $E_u'$, we calculate it by multiplying the expected infections $E_u$ with the cumulative probability of infection occurring after $T$.

\subsection{H2-MARL human mobility restriction strategy model}
\label{section:H2-MARL_strategy_model}

\begin{figure}
	\centering
	\includegraphics[width=\linewidth]{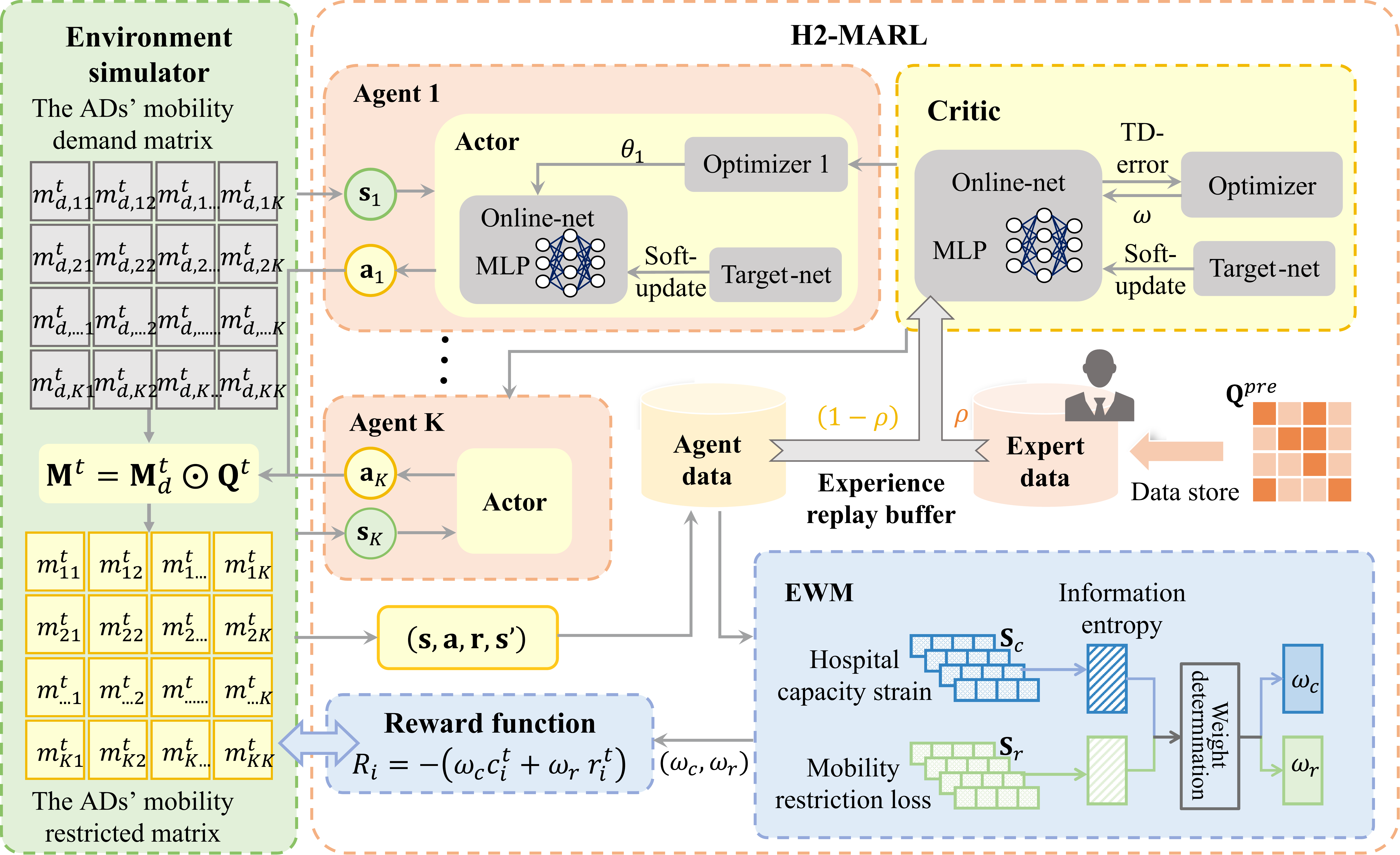}
	\caption{Overview of H2-MARL model.}
	\label{fig:H2-MARL_framework}
\end{figure}

H2-MARL is a multi-agent mobility restriction strategy model designed to generate the optimal mobility quota $\mathbf{Q}^t$ in (\ref{eq:Mt_Qt}) for restricting human mobility between ADs, achieving a balance between hospital capacity strain and mobility restriction loss. An overview of H2-MARL is shown in Fig. \ref{fig:H2-MARL_framework}. H2-MARL treats each AD as an agent, each with an actor network that receives its own observation from the environment simulator and outputs the action. The simulator then provides feedback rewards and the next state for state-action pairs based on the reward function. To adapt to environmental changes, the reward function employs the entropy weight method (EWM) to adjust the weights of the dual objectives dynamically. The model utilizes a shared critic network and randomly samples trajectories of agent-environment interactions in proportion from two experience replay buffers, which store agent and expert data, respectively. The critic network evaluates the actions of all agents and performs soft-update for the actor and critic networks, maximizing the overall reward of the environment.


\subsubsection{Dual-objective decision problem formulation}
\label{section:Q_calculation}
The dual objectives of the H2-MARL strategy are to minimize hospital capacity strain as well as human mobility restriction loss. Hospital capacity strain is caused by the increase in hospitalizations $H_i^t$ of each AD, and mobility restriction loss is caused by the restriction on mobility quota $\mathbf{q}_i^t$ for each AD.

For agent $i$ on day $t$, we define the hospital capacity strain index (defined as $c_{i}^{t}$) and the mobility restriction loss index (defined as $r_i^{t}$), calculated respectively by the $\phi$ and $\psi$ functions as follows:
\begin{equation}
    c_i^t = \psi (H_i^t), \quad r_i^t = \phi (\mathbf{M}_d^t, \mathbf{q}_i^t),
\end{equation}
where $\phi(\cdot)$ and $\psi(\cdot)$ are defined in (\ref{eq: U_i^t_function}), (\ref{eq: C_i^t}). Then, the dual-objective decision problem can be formulated as:
\begin{equation}
  \mathbf{q}_i^t = \arg \max_{\mathbf{q}_i^t} {\sum_{\tau=t}^T \xi (c_i^\tau, r_i^\tau)},
  \label{eq: Q^t_objective}
\end{equation}
where $\xi(\cdot)$ represents the overall objective function and is utilized as the reward function of H2-MARL in (\ref{eq: reward_function_R}). $\xi$ satisfies ${\partial^2 \xi}/{\partial c_i \partial r_i} < 0$ because of the trade-off nature between hospital capacity and mobility retaining. Additionally, $\xi$ should meet some practical requirements, i.e., $H_i^t$ is below the actual maximum capacity of AD.

\subsubsection{Dual-objective MDP}
The restriction decision-making problem can be naturally formalized as a dual-objective Markov decision process (MDP), represented by a tuple $(\mathcal{S}, \mathcal{A}, Pr, \mathcal{R}, \eta)$. $Pr$ is the state transition function, and $\eta$ is a discount factor, with $\eta \in [0, 1]$. $\mathcal{S}$, $\mathcal{A}$, and $\mathcal{R}$ represent the state space, action space, and reward function, respectively, which are detailed as follows.

{\bf{State space $\mathcal{S}$}}. $\mathcal{S}$ indicates the state space of all agents, represented by the direct product of each agent's state space. $\mathcal{S}$ and the state of each agent are defined as
\begin{equation}
    \begin{aligned}
    \mathcal{S} &= \mathcal{S}_1 \times \mathcal{S}_2 \times \ldots \times \mathcal{S}_K,\\
    \mathbf{s}_i &= (P_i, \mathbf{D}^t_i, \Delta \mathbf{D}^t_i, L_i^t, \bar{\mathbf{M}}_{d,i}^t), \quad \mathbf{s}_i \in \mathcal{S}_i, \\
    \end{aligned}
\end{equation}
where $\mathcal{S}_i$ is the state space of agent $i$. $\mathbf{s}_i$ represents the state of agent $i$ on day $t$, defined as a tuple, and $P_i$ denotes the encoded position for agent $i$. The vector $\Delta \mathbf{D}^t_i$, defined as $\Delta \mathbf{D}^t_i = [\Delta S^t_i, \; \Delta I^t_i, \; \Delta H^t_i, \; \Delta R^t_i]$, indicates the change in $\mathbf{D}^t_i$ within a unit of time. $\bar{\mathbf{M}}_{d, i}^t$ represents the average of the mobility demand for agent $i$ on day $t$, which is given by
\begin{equation}
    \bar{\mathbf{M}}_{d, i}^t = \frac{\sum_{j=1}^K m_{d, ij}^t}{K},
\end{equation}
$L_i^t$ records the accumulated restriction loss expressed as
\begin{equation}
    L_i^t = \sum_{\tau = 0}^{t-1} \lambda^{t-\tau} \frac{\sum_{j=1}^K(m_{d, ij}^\tau - m_{ij}^\tau)}{\bar {\mathbf{M}}^{\tau}_{d, i}},
    \label{eq:L_i^t}
\end{equation}
where $L_i^t$ is the accumulated loss due to mobility restriction from the initial time to the current day $t$. We use a decay factor $\lambda$ to represent the decay rate of the influence over time.

For each agent, it is not feasible to access all global information, and most of the information is not necessary for decision-making. In our setup, only information related to the agent itself is retained to represent the environment's state, reducing the dimensionality of the state space.

{\bf{Action space $\mathcal{A}$}}. $\mathcal{A}$ is the Cartesian product of the action spaces of all agents. Each agent's action represents the mobility restriction that determines the mobility quota from each AD to other ADs on day $t$. Therefore, we define $\mathcal{A}$ in the following:
\begin{equation}
    \begin{aligned}
    \mathcal{A} &= \mathcal{A}_1 \times \mathcal{A}_2 \times \cdots \times \mathcal{A}_K, \\
    \mathbf{a}_i &= \mathbf{q}_i^t, \quad \mathbf{a}_i \in \mathcal{A}_i, \\
    \end{aligned}
\end{equation}
where $\mathcal{A}_i$ denotes the action space of agent $i$ and $\mathbf{a}_i$ represents the action vector of agent $i$ on day $t$.

{\bf{Reward function $\mathcal{R}$}}. The reward function is defined as $\mathcal{R} = (\mathcal{R}_1, \mathcal{R}_2, \ldots, \mathcal{R}_K)$, where $\mathcal{R}_i$ represents the immediate reward of agent $i$ in state $\mathbf{s}_i$ based on its action $\mathbf{a}_i$. $\mathcal{R}_i$ is designed to represent the objective function $\xi$ in (\ref{eq: Q^t_objective}) and is formulated as 
\begin{equation}
    \mathcal{R}_i = -(\omega_c c_i^t + \omega_r r_i^t),
    \label{eq: reward_function_R}
\end{equation}
where the weights $\omega_c$ and $\omega_r$ represent the trade-off between the two indices, with $\omega_c + \omega_r = 1$. We denote the hospital capacity index $c_i^t=\psi (H_i^t)$ as follows:
\begin{equation}
  c_i^t = k_i \exp{(\frac{H_i^t}{h_0})},
  \label{eq: U_i^t_function}
\end{equation}
where the hyperparameter $k_i$ represents the initial level of hospital capacity in AD $i$. $h_0$ is also a hyperparameter, determining the rate of increase in hospital capacity strain with the number of hospitalizations. As $H_i^t$ increases, the hospital capacity strain $c_i^t$ will grow exponentially.

The human mobility restriction loss index $r_i^t = \phi (\mathbf{M}_d^t, \mathbf{q}_i^t)$ is designed as follows:

\begin{equation}
    r_i^t = \exp{(\frac{L_i^t}{l_0})} \frac{\sum_{j=1}^{K}{(m_{d ,ij}^t - m_{ij}^t)}}{\bar{\mathbf{M}}_{d, i}^t},
    \label{eq: C_i^t}
\end{equation}
where the hyperparameter $l_0$ determines how large the penalty is for continuously limiting the same AD. Eq. (\ref{eq: C_i^t}) represents the total loss cost resulting from mobility restriction in AD $i$. Considering the continuing impact of past mobility restrictions on the current state, we utilize an exponential function to illustrate the decay over time.

The weights of objectives, $\omega_c$, and $\omega_r$, directly affect the reward outcomes in (\ref{eq: reward_function_R}), thereby guiding the direction of optimization. The weights reflect the relative importance and priority among objectives in the current decision-making problem, which should be dynamically adjusted in response to changes in environmental preferences. To obtain $\omega_c$ and $\omega_c$, the entropy weight method \citep{zhu2020effectiveness} measures the uncertainty of each objective based on information entropy, providing an objective value for the weights. $\omega_c$ and $\omega_r$ are given as
\begin{equation}
    \label{eq:weight_entropy}
        w_c = \frac{1 - H_c}{2 - H_c - H_r}, \quad
        w_r = \frac{1 - H_r}{2 - H_c - H_r},
\end{equation}
where
\begin{equation}
    \begin{aligned}
        H_c &= -k\sum_{t=0}^T \mathbf{P}_c^t \ln \mathbf{P}_c^t, \;\; H_r = -k \sum_{t=0}^T \mathbf{P}_r^t \ln \mathbf{P}_r^t, \\
        \mathbf{P}_c^t &= \frac{\mathbf{S}_c^t}{\sum^T_{t=0}\mathbf{S}_c^t}, \;\; \mathbf{P}_r^t = \frac{\mathbf{S}_r^t}{\sum^T_{t=0}\mathbf{S}_r^t}.
    \end{aligned}
\end{equation}
Here we sample data generated by objective functions (\ref{eq: U_i^t_function}) and (\ref{eq: C_i^t}) to obtain normalized time series matrices for $T$ days: $\mathbf{S}_c = [\mathbf{S}_c^t]_{t=0}^T$ and $\mathbf{S}_r = [\mathbf{S}_r^t]_{t=0}^T$, where $\mathbf{S}_c^t=[c_i^t]_{i=1}^K$ and $\mathbf{S}_r^t=[r_i^t]_{i=1}^K$. Both $\mathbf{S}_c$ and $\mathbf{S}_r$ are treated as negative indicators. $\mathbf{P}_c^t$ and $\mathbf{P}_r^t$ represent the proportion of the normalized sample $\mathbf{S}_c$ and $\mathbf{S}_r$, respectively. $H_c$ and $H_r$ indicate the information entropy for the $\mathbf{S}_c$ and $\mathbf{S}_r$. The constant $k=\frac{1}{\ln T}$. Therefore, the greater the information entropy in (\ref{eq:weight_entropy}) of the objective, the more uniform the information distribution, which indicates its relatively lower importance, and its weight will be smaller.

\subsubsection{Policy optimization}
The policy optimization process for H2-MARL is shown in Fig. \ref{fig:H2-MARL_framework}. The model consists of $K$ agents, each utilizing a deterministic policy actor network to generate restriction policy $\mu_i(\mathbf{s}_i; \theta_i)$ w.r.t. parameter $\theta_i$(abbreviated as $\mu_i$). Besides, all agents in the model utilize a shared critic network that provides the estimated value $Q(\mathbf{s}, \mathbf{a}; \omega)$ for state-action pairs. $\mathbf{s} = [\mathbf{s}_1, \mathbf{s}_2, \ldots, \mathbf{s}_K]$ indicates the global state information, and $\mathbf{a} = [\mathbf{a}_1, \mathbf{a}_2, \ldots, \mathbf{a}_K]$ denotes the set of actions taken by all agents. The shared critic network facilitates inter-agent collaboration and information sharing while solving the curse of dimensionality.

H2-MARL has two replay buffers, one is an agent experience replay buffer, which stores the trajectories of all agents interacting with the environment. The trajectory is represented as a tuple ($\mathbf{s}$, $\mathbf{a}$, $\mathbf{r}$, $\mathbf{s'}$), where $\mathbf{r}=[r_i]_{i=1}^K$ denotes the joint rewards for all agents, with $r_i = \mathcal{R}_i$, and $\mathbf{s'}$ indicates the next state. The other one is an expert experience replay buffer, which is filled with trajectory tuple data generated by a preset expert policy. The expert policy, defined as $\mathbf{Q}^{pre}$, with $\mathbf{Q}^{pre} = [q^{pre}_{ij}]_{1\leq i, j\leq K}$, is given by
\begin{equation}
    q_{i,j}^{pre}=\begin{cases}0&\quad\text{if} \ \  H_i^t > X_h \ \ \text{and} \ \ L_i^t < X_l\\1&\quad\text{else.}\end{cases}
    \label{eq: expert_strategy_q}
\end{equation}
where $X_h$ and $X_l$ denote the thresholds for the number of hospitalizations and the accumulated restriction loss. The expert policy will fully lockdown an AD if two conditions are met: first, the number of hospitalizations $H_i^t$ exceeds the threshold $X_h$; second, the accumulated restriction loss $L_i^t$ of the AD does not exceed the threshold $X_l$.

The critic randomly samples batches of data from both the agent and expert experience replay buffers simultaneously. This ensures stable convergence of the learning process while satisfying the algorithm's exploration requirements. The hyperparameter $\rho$ in Fig. \ref{fig:H2-MARL_framework} represents the sampling ratio, which controls the proportion of data sourced from the expert policy compared to the agent's own experience. The loss function $L(\omega)$ of the shared critic is defined as follows:
\begin{equation}
    \mathcal{L}(\omega) = \mathbb{E}_{(\mathbf{s}, \mathbf{a}, \mathbf{r}, \mathbf{s'})}[(Q(\mathbf{s}, \mathbf{a}) - y)^2], 
\end{equation}
\begin{equation}
    y = \frac{1}{K}\sum_{i=1}^K{r_i} + \eta \cdot Q(\mathbf{s'}, \mathbf{a'})\big|_{\mathbf{a}_i' = \mu'_i(\mathbf{s}_i')},
\end{equation}
where $\mu' = \{\mu_{\theta_1'}, \mu_{\theta_2'}, \ldots, \mu_{\theta_K'}\}$ is the set of target policies with delayed parameters $\theta_i'$.

\section{Experiments}
\label{section:experiments}
In this section, we construct the experiment datasets and evaluate the performance of our proposed environment simulator and H2-MARL strategy model in four representative cities of varying scales. Furthermore, we conduct ablation studies to assess the effectiveness of our proposed methods for updating infection rates, adjusting reward weights, and enhancing expert knowledge.

\subsection{Datasets}

\begin{table}[htbp]
  \caption{Sample of resident travel trajectory data.}\label{tab:sample_data}
  \centering
  \resizebox{\textwidth}{!}{%
  \begin{tabular}{cccccccc}
  \toprule
  \textbf{Uid} & \textbf{Start\_time} & \textbf{Start\_lng} & \textbf{Start\_lat} & \textbf{End\_time} & \textbf{End\_lng} & \textbf{End\_lat} & \textbf{Weight} \\
  \midrule
  73681 & 10/01/2021, 7:17:54 & 113.24453 & 23.12547 & 10/01/2021, 7:44:36 & 113.25449 & 23.12538 & 5.65490 \\
  72548& 10/01/2021, 9:23:58 & 113.24643 & 23.11732 & 10/01/2021, 10:26:10 & 113.26237 & 23.11932 & 5.47417 \\
  60508 & 10/02/2021, 8:50:49 & 113.23547 & 23.13281 & 10/02/2021, 9:53:57 & 113.24598 & 23.12638 & 4.98295 \\
  52467 & 10/03/2021, 11:55:15 & 113.26219 & 23.14232 & 10/03/2021, 12:43:31 & 113.24815 & 23.15147 & 5.01604 \\
  58961 & 10/03/2021, 10:11:43 & 113.25558 & 23.12566 & 10/03/2021, 11:26:10 & 113.24906 & 23.11252 & 5.22890 \\
  64565 & 10/04/2021, 12:24:41 & 113.21075 & 23.14612 & 10/04/2021, 12:56:12 & 113.22288 & 23.12221 & 6.79713 \\
  85465 & 10/05/2021, 15:51:35 & 113.27187 & 23.11368 & 10/05/2021, 16:00:21 & 113.26470 & 23.10409 & 5.01604 \\
  77451 & 10/05/2021, 22:27:08 & 113.28418 & 23.12420 & 10/05/2021, 22:56:13 & 113.27025 & 23.12119 & 5.01605 \\
  65311 & 10/06/2021, 3:46:12 & 113.26459 & 23.10322 & 10/06/2021, 4:19:22 & 113.25864 & 23.10929 & 5.01605 \\
  73124 & 10/06/2021, 21:13:54 & 113.28290 & 23.14136 & 10/06/2021, 21:41:28 & 113.27262 & 23.14857 & 5.01605 \\
  \bottomrule
  \end{tabular}}
\end{table}

\begin{table}[htbp]
  \caption{Dataset statistics}\label{tab:dataset_statistics}
  \centering
  \resizebox{\textwidth}{!}{%
  \begin{tabular}{ccccccc}
  \toprule
  \textbf{City} & \textbf{Area ($\mathbf{km^2}$)} & \textbf{Population} & \textbf{Duration (day)} & \textbf{Records} & \textbf{ADs} & \textbf{Administrative levels} \\
  \midrule
  Guangzhou & 7,434.40 & 9,677,188 & Oct. 1, 2021-Oct. 31, 2021 & 495,261,517 & 167 & provincial capital \\
  Wuxi & 4627.47 & 1,888,094 & June. 1, 2023-June. 30, 2023 & 306,984,995 & 92 & prefecture-level \\
  Chongqing & \multirow{2}*{4779.00} & \multirow{2}*{3,518,314} & \multirow{2}*{June. 1, 2023-June. 30, 2023} & \multirow{2}*{377,429,086} & \multirow{2}*{156} & \multirow{2}*{municipality} \\
  (center district) & ~ & ~ & ~ & ~ & ~ & ~ \\
  Ezhou & 1596.46 & 1,078,668 & June. 1, 2023-June. 30, 2023 & 40,281,784 & 27 & prefecture-level \\
  \bottomrule
  \end{tabular}}
\end{table}

We construct a COVID-19 infection historical dataset for reference purposes and a human mobility OD dataset for modeling mobility demand.

\subsubsection{COVID-19 infection dataset}
The dataset covers the COVID-19 situation in four cities in China, specifically including Guangzhou (from Oct. 1, 2022, to Oct. 30, 2022) $\footnote{data source: Guangzhou Municipal Health Commission.}$, Chongqing (from Nov. 1, 2022, to Dec. 1, 2022) $\footnote{data source: Chongqing Municipal Health Commission.}$, Jiangsu (from Nov. 10, 2022, to Dec. 11, 2022) $\footnote{data source: Jiangsu Commission of Health.}$, and Hubei (from Nov. 5, 2022, to Dec. 7, 2022) $\footnote{data source: Health Commission of Hubei Province.}$. The dataset contains four fields: the current number of infectors, the daily number of new infectors, the number of hospitalizations, and the number of discharged patients.

\subsubsection{Human mobility OD dataset}
Existing OD datasets are insufficient for township-level ADs that encompass cities of various scales. To address this deficiency, we construct an OD dataset based on mobile signaling data and summarize the detailed information of the dataset in Table~\ref{tab:dataset_statistics}. The dataset contains over 1 billion real-world historical population trajectories, covering 3 administrative levels of cities: provincial capitals, prefectural-level cities, and central districts of municipalities, with population sizes ranging from millions to tens of millions.

The trajectory data is obtained from the mobile phone signaling data (desensitized data) on the Unicom Smart Footprint platform. Resident travel trajectory data from Guangzhou, the center district of Chongqing, Wuxi (in Jiangsu Province), and Ezhou (in Hubei Province) in China is extracted. The data contains eight fields: uid, start\_time, start\_lng, start\_lat, end\_time, end\_lng, end\_lat and weight. Specifically, uid is the user ID, start\_time is the origin traffic point departure time, start\_lng and start\_lat are the longitude and latitude coordinates of the origin traffic point, end\_time is the destination traffic point departure time, end\_lng and end\_lat are the longitude and latitude coordinates of the destination traffic point, and weight is the number of trips between the origin and destination traffic points (extrapolation of the number of the Unicom subscribers to the full number of people). Table~\ref{tab:sample_data} shows the sample data of the travel trajectory data.

The process of obtaining trajectory data is as follows:
\begin{enumerate}
    \item{Upload the city's township-level layer boundary file in CSV format, where the spatial field is represented in WKT format.}
    \item{Establish the spatial relationship between travel trajectory data and the city boundary grid.}
    \item{Calculate the cross-regional mobility OD matrix by time period.}
\end{enumerate}

\subsection{Validation of the D-SIHR model}
\subsubsection{Evaluation metric and hyperparameter settings}

We utilize the goodness-of-fit parameter $R^2$ \citep{nystrom2019impact} as the evaluation metric for the D-SIHR infection model, which is defined as
\begin{equation}
    R^2 = 1 - \frac{\sum_{t=0}^T{(I-\hat{I_t})}^2}{\sum_{t=0}^T{(I-\bar{I_t})}^2},
    \label{R^2_para}
\end{equation}
where $\hat{I_t}$ and $I_t$ represent the estimated and actual values of the model on day $t$, respectively, and $\bar{I_t}$ denotes the average number of infections. $T$ indicates the total time. Eq. (\ref{R^2_para}) shows that the closer $R^2$ is to 1, the stronger the model's ability to characterize the target variable, indicating better simulation performance.

The hyperparameter settings for the D-SIHR are presented in Table~\ref{tab:appendix_D-SIHR_para}. According to the research in \citep{li2020early} the COVID-19  serial interval time approximation obeys the gamma distribution of the shape parameters $\alpha$=7.5 and $\beta$ = 3.4.

\begin{table}[htbp]
  \caption{The hyperparameters of the D-SIHR model experiment.}\label{tab:appendix_D-SIHR_para}
  \centering
  \resizebox{0.6\textwidth}{!}{%
  \begin{tabular}{cccccc}
  \toprule
  city & $\beta_*^{t} (t=0)$ & $R_{t} (t=0)$ & $\gamma$ & $\theta$ & $\delta$ \\
  \midrule
  Guangzhou & 0.47 & 2.10 &0.0096 & 0.13 & 0.19 \\
  Chongqing & \multirow{2}*{0.36} & \multirow{2}*{1.60} & \multirow{2}*{0.0096} & \multirow{2}*{0.11} & \multirow{2}*{0.16} \\
  (center district) & ~ & ~ & ~ & ~ \\
  Jiangsu & 0.51 & 2.30 & 0.0046 & 0.13 & 0.18  \\
  Hubei & 0.47 & 2.10 & 0.0096 & 0.10 & 0.13\\
  \bottomrule
  \end{tabular}}
\end{table}

\subsubsection{Results and analysis}
\begin{figure}
	\centering
	\includegraphics[width=0.8\linewidth]{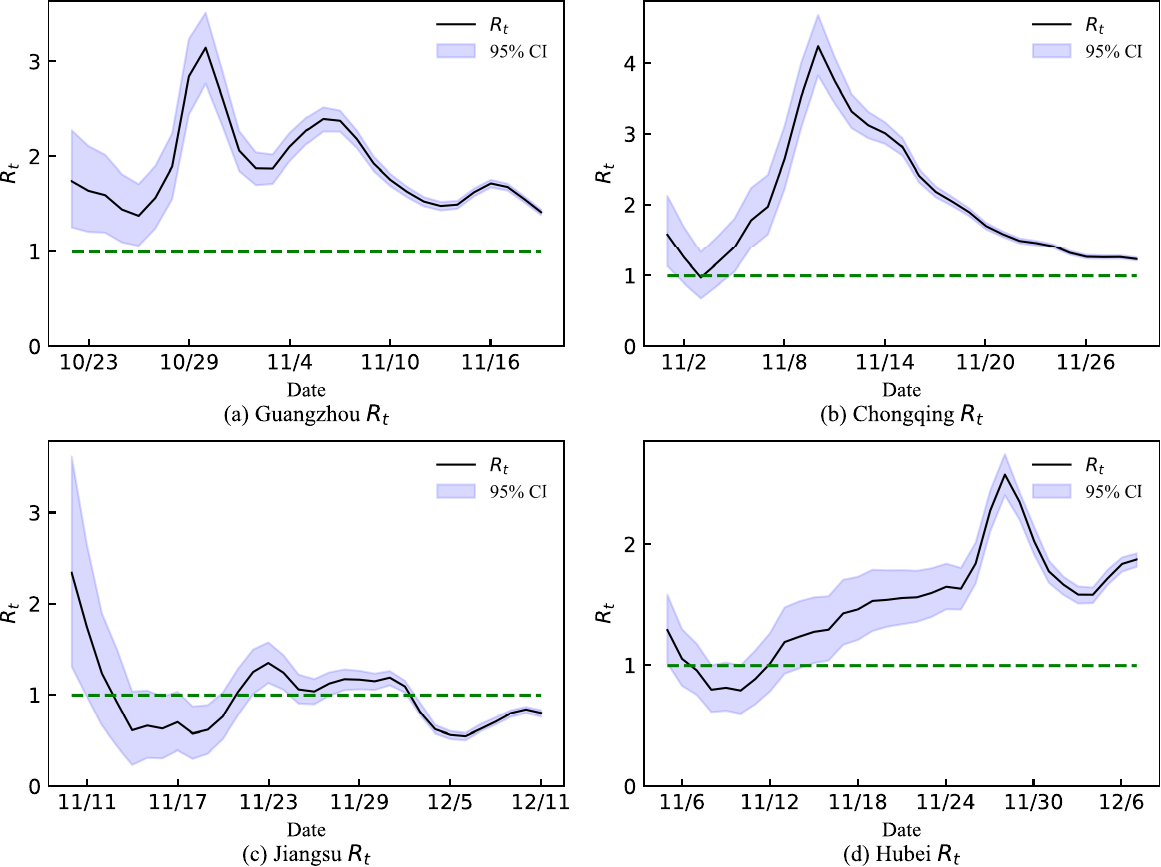}  
	\caption{Simulation curves of the effective reproduction number ($R_t$) for COVID-19 and the 95\% confidence intervals (CI): Guangzhou (a), Chongqing (b), Jiangsu (c), and Hubei (d). The blue shaded area represents the 95\% confidence interval, and the green dashed line highlights the scenario where the epidemic is under control ($R_t$ = 1).}
	\label{fig:gz_cq_rt_result}
\end{figure}

We calculate the effective reproduction number $R_t$ for four cities over the corresponding time periods. As shown in Fig. \ref{fig:gz_cq_rt_result}, the results indicate that all four cities experienced severe outbreaks initially, but the transmission dynamics varied across cities. $R_t$ of the COVID-19 outbreak in Guangzhou fluctuated several times, peaking on the ninth day (Oct. 30) at $R_t = 3.144$, and then showed a declining trend, gradually approaching 1. In Chongqing, $R_t$ rose rapidly in the early stages, peaking on the tenth day (Nov. 10) at $R_t = 4.239$, after which it began to decline rapidly, eventually approaching 1. In Jiangsu, $R_t$ decreased from its peak at $R_t = 2.345$ and fell below 1 on Nov. 13. Although there were minor rebounds and fluctuations thereafter, it remained around 1 overall. In Hubei, $R_t$ initially fell below 1, but then rose sharply starting on the fifth day (Nov. 9), reaching a maximum value of $R_t = 2.579$ on Nov. 28, followed by a temporary decline and subsequent increase.

Changes in $R_t$ reflect the impact of epidemic restriction strategies on infection capacity. Values of $R_t$ greater than 1 indicate that the epidemic will continue to spread, while $R_t < 1$ signifies that the infection is under control \citep{nishiura2009effective}. The results from the four cities demonstrate that $R_t$ exhibits significant fluctuations influenced by restriction measures during the transmission process and eventually shows a downward trend.

\begin{figure}
	\centering
	\includegraphics[width=0.8\linewidth]{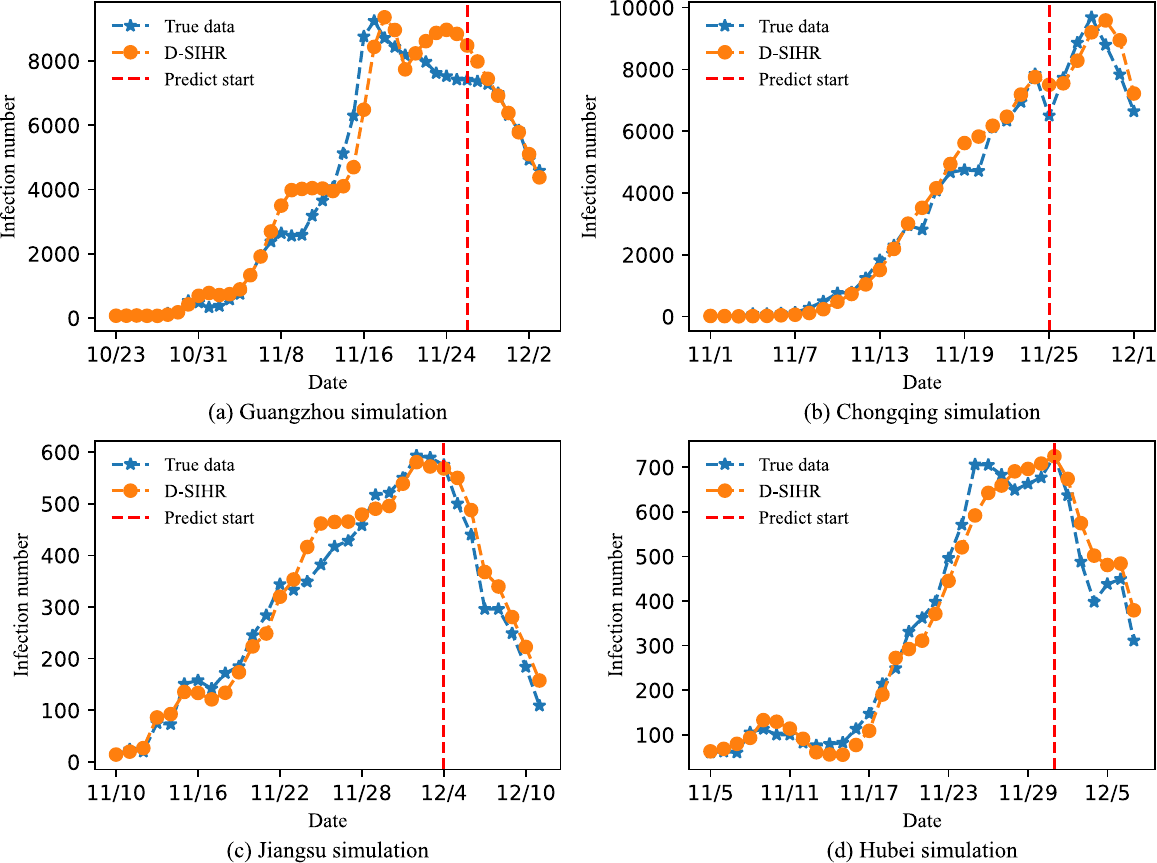}  
	\caption{The results from Guangzhou (a), Chongqing (b), Jiangsu (c), and Hubei (d) in China. The orange solid line represents the model-simulated daily infections, the blue dashed line corresponds to the historical data published by authoritative agencies, and the red dashed line shows the predicted results for the next seven days.}
	\label{fig:gz_cq_simulation_result}
\end{figure}

For each city, we utilize the D-SIHR to simulate the infection process and predict the daily number of infectors for the next 7 days. As shown in Fig. \ref{fig:gz_cq_simulation_result}, predicted infection numbers by the proposed environment simulator align closely with the actual infection numbers, effectively capturing critical trends such as the peak infection counts and subsequent declines.

Specifically, for Guangzhou, the D-SIHR model accurately simulates the changes in infection dynamics. However, a slight underestimation of the decline in infections occurs. In Chongqing, the predicted results are closely consistent with actual data, though the simulated variations appear smoother compared to the real-world data. For Jiangsu, the D-SIHR performs well, with only a minor overestimation toward the end of the prediction period. In Hubei, it accurately replicates the infection peak and subsequent changes. The $R^2$ metrics of the D-SIHR model for Guangzhou, Chongqing, Jiangsu, and Hubei are 0.9444, 0.9663, 0.9328, and 0.9787, respectively, indicating a good fit to the result. D-SIHR model demonstrates strong simulation and predictive capabilities, though the trend variations tend to be slightly smoother than observed. This might be due to the model's parameter estimation or specific assumptions not fully capturing certain abrupt factors, such as policy changes. Nevertheless, it does not significantly affect the model's overall trend assessment.

\subsection{Evaluation of the H2-MARL model}
In this section, we compare the performance of our proposed H2-MARL model with several benchmark strategies for suppressing the spread of infectious diseases, utilizing the D-SIHR model as the epidemic environment model.

\subsubsection{Baselines and metrics}
To evaluate the effectiveness of H2-MARL, we set five baseline models compared with H2-MARL. We provide a brief explanation of these models below.

\begin{itemize}
  \item{\bf{No policy}}. It refers to the epidemic progressing naturally without implementing any interventions on human mobility at any decision point.
  \item{\bf{Count threshold-based policy $\pi^T$}.} A heuristic strategy model in \citep{lin2010optimal} is adopted, which sets the mobility ratio $q_{ij}^t$=10\% when the number of new cases $\Delta I_i^t$ in the AD exceeds $N_i^t/1000$, and sets the mobility ratio $q_{ij}^t$=100\% when the number of new cases $\Delta I_i^t<3$.
  \item{\bf{Occurrence-based mitigation policy $\pi^M$}.} Based on a heuristic strategy model in \citep{merl2009statistical}, if the AD records new infections totaling $\sum_{\tau=0}^{t=6}\Delta I_i^\tau > N_i / 1000$ over the past 7 days, the mobility ratio is set to $q_{ij}^t = 50\%$. Otherwise, there are no restrictions on population movement.
  \item{\bf{Occurrence-based suppression policy $\pi^S$}.} A heuristic strategy model in \citep{ludkovski2010optimal} is employed. For ADs where new total infections $\sum_{\tau=0}^{t=6} \Delta I_i^{\tau} \geq 1$ within the first 7 days, the mobility rate is set to $q_{ij}^t = 30\%$. After 7 days, the rate is adjusted to $q_{ij}^t = 10\%$. If no new infections occur within 7 days($\sum_{\tau=0}^{t=6}\Delta I_i^{\tau} < 1$), the rate is set to $q_{ij}^t = 50\%$, and after 14 days, it increases to $q_{ij}^t = 90\%$.
  \item {\bf{Reinforcement learning-based mitigation policy $\pi^M_{RL}$}}. Based on an adaptive strategy model in \citep{song2020reinforced}, the city is regarded as an agent, and an RL-based graph neural network (GNN) is adopted. This agent dynamically adjusts the human mobility quota $\mathbf{Q^t}$ based on the current feedback from the environment, generating an optimal policy to maximize the cumulative reward of the environment.
\end{itemize}

The strategies $\pi^T$, $\pi^M$, and $\pi^S$ are based on observations from real life and have been frequently considered in recent literature \citep{wan2021multi}.

To evaluate the performance of baseline and H2-MARL models in terms of multi-objective trade-off capabilities, we introduce a dual-objective optimization metric\citep{riquelme2015performance, yan2007diversity}, defined as $D$. Specifically, we set the optimal values of the two objectives as the ideal solution, and then calculate the Euclidean distance between the solution of the mobility restriction strategy model and the ideal solution. The smaller the distance, the closer it is to the ideal solution, indicating better Pareto optimality of the strategy. The metric is given by

\begin{equation}
  D = \sqrt{\left( \frac{c - c^{\text{min}}}{c^{\text{max}} - c^{\text{min}}} - \dot{c} \right)^2 + \left( \frac{r - r^{\text{min}}}{r^{\text{max}} - r^{\text{min}}} - \dot{r} \right)^2}
\end{equation}
where $c^{max}$, $c^{min}$, $r^{max}$ and $r^{min}$ represent the minimum and maximum values of the objectives hospital capacity strain $c$ and mobility restriction loss $r$, respectively. $\dot{c}$ and $\dot{r}$ denote the normalized objective values of the ideal solution (set $\dot{c} = 0, \dot{r}=0$).

\begin{figure}
	\centering
	\includegraphics[width=\linewidth]{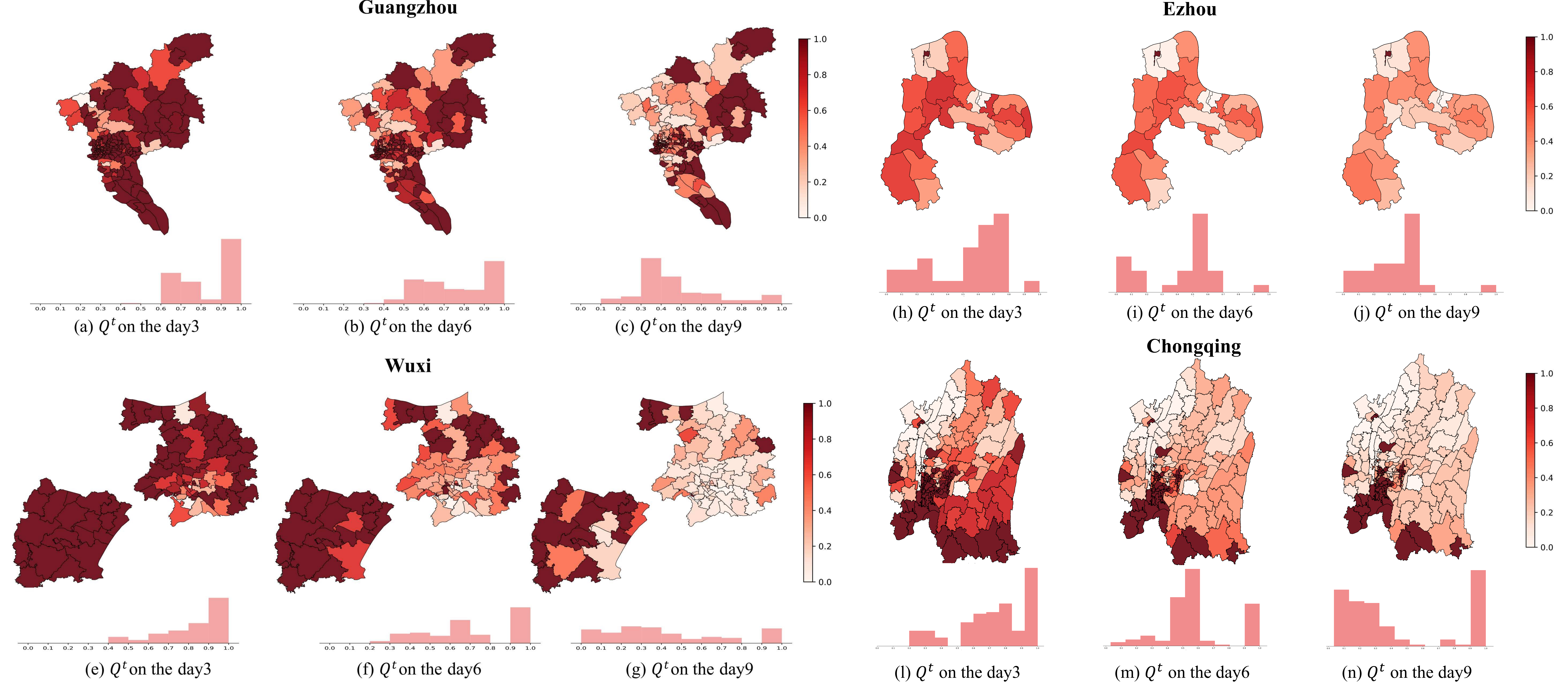}  
	\caption{The spatial distribution and histograms of $Q^t$ for four cities, (a-c) Guangzhou, (e-g) Wuxi, (h-j) Ezhou, and (i-n) Chongqing(center district), on days 3, 9, and 12, respectively. The color bar from dark to light represents the human mobility quota from completely free $(q_{ij}^t=100\%)$ to completely locked down $( q_{ij}^t=0\%)$. Each histogram summarizes the distribution of quota rates in the respective period.}
	\label{fig:4_cities_H2-MARL_result}
\end{figure}

\subsubsection{Constraints and success criteria in strategy exploration}
The success criterion for the mobility restriction strategy is designed such that the number of new infections in all ADs is less than 1, which is regarded as an effective achievement of zero new cases. Additionally, during the control process, the current number of hospitalizations does not exceed 6.92 per 1,000 of the total population in each AD $\footnote{The data is sourced from the ``Statistical Bulletin on the Development of Health Care in China in 2022'' by the National Health Commission of the People's Republic of China, specifically regarding the number of hospital beds per thousand population.} $.

The dual-objective trade-off and the exponentially growth nature of epidemics complicate agent learning and increase the risk of convergence to locally optimal policies. We designed 3 constraint rules for action space pruning based on heuristic rules to address this problem by preventing agents from exploring evidently unreasonable policies.

\begin{itemize}
  \item {\bf{Time-out threshold}}: the current time step reaches the set maximum time threshold.
  \item {\bf{Hospitalization threshold}}: the number of ADs where $H_i^t$ exceeds the hospitalization threshold $H_T$ is more than 20\% of the total number of ADs $K$, i.e., $\sum_{i=1}^{K} \mathbb{I}(H_i^t > H_T) > 0.2K$, where $\mathbb{I}(\cdot)$ is set as an indicator function.
  \item {\bf{Lockdown threshold}}: the number of ADs where $L_i^t$ exceeds the hospitalization threshold $L_T$ is more than 20\% of the total number of ADs $K$, i.e., $\sum_{i=1}^{K} \mathbb{I}(L_i^t > L_T) > 0.2K$.
\end{itemize}

If the environmental feedback from the agent's strategy meets the predefined rules, we immediately halt exploration, terminate the current episode, and impose a substantial penalty to enhance the efficiency of the learning process.

\subsubsection{Results and analysis}
We set the experimental parameters for the strategy optimization process: $\lambda = 0.99$, $l_0= 64$, $h_0=72$, $k_i = 0.8, i \in [1, \ldots, K]$. For the thresholds of expert policy in Eq. (\ref{eq: expert_strategy_q}), we set $X_h = 100$, $X_l = 168$. For the strategy constraints, we set the time threshold as 60, the hospitalization threshold as $H_T = 6.92*N/1000$, where $N$ is the total number of population in the AD region, and the lockdown threshold as $L_T = 336$. During the training process, we set the learning rate to 0.0001, $\eta = 0.9$, and $\rho = 0.5$, decreasing by 0.1 every 200 steps until reaching 0. The maximum number of steps per episode to 300. Considering the randomness of training, we train H2-MARL with different random seeds 10 times for each set of configurations and report the means and standard deviations of all experiment metrics as the result.

The proposed H2-MARL human mobility restriction strategies are applied in four urban epidemic scenarios. The quota rates generated by these strategies in each township-level AD are shown in Fig. \ref{fig:4_cities_H2-MARL_result}. It visualizes the mobility quotas $q_{{i_0}j}$ from the initial infection outbreak AD $i_0$ to other ADs on days 3, 6, and 9 after the outbreak in four cities. Variations in color depth indicate the quota for each AD is adjusted automatically by the H2-MARL strategy, where deeper color means more relaxed mobility restriction. As the infection spreads with time, $q_{{i_0}j}$ generally decreases, indicating stricter restrictions. In addition, the H2-MARL strategy tends to allocate more quotas during periods of low risk and low mobility while allocating fewer quotas during periods of high risk or high mobility. Moreover, the spatial distribution of the results indicates that the strategies are related to the population size of the ADs, with the strategies tending to allocate higher mobility to sparsely populated ADs.

We compared H2-MARL with other baseline models in four cities to compare and analyze the strategy of each model, as shown in Table~\ref{tab:comparison_gz_wx_cq_ez}. To evaluate the effectiveness of the model strategies, we utilized metrics including the average daily number of hospitalizations of all ADs $\bar{H}$, the average daily mobility quota of all ADs $\bar{\mathbf{Q}}$, the average hospital capacity strain index $\bar{c}$, the average mobility restriction loss index $\bar{r}$, time to reach the success (TTS) and $D$. The TTS denotes the time from the initiation of the strategy to the achievement of the restriction success criteria, measured in days. The direction of the arrows for the metrics indicates the expected trend, where $\downarrow$ signifies that lower values are preferable, while $\uparrow$ indicates that higher values are better. If the target was not achieved within the maximum training time, it is denoted by $/$.

The tabular results indicate that our proposed and baseline models exhibit epidemic control capabilities relative to the no-policy model. Specifically, the count threshold-based policy $\pi^T$ demonstrates outstanding performance in reducing hospitalization numbers $\bar{H}$. However, $\pi^T$ shows relatively weak performance in restricting human mobility, where the average mobility quotas $\bar{\mathbf{Q}}$ are as low as 0.21 and 0.27 in Wuxi and Chongqing, respectively, indicating overly stringent restrictions on movement. Moreover, in the large-scale cities, i.e., Guangzhou and Wuxi, $\pi^T$ fails to achieve restriction success within the specified timeframe, with missing TTS values indicated by $/$, signifying an unacceptable risk of restriction failure.

In contrast, $\pi^M$ is more lenient in terms of mobility restrictions with higher mobility quotas compared to $\pi^T$. Besides, alleviating infections leads to significantly higher $\bar{H}$ values for Wuxi and Chongqing, thereby negatively impacting its epidemic control performance (indicated by TTS), particularly in Chongqing and Ezhou as TTS=$/$. Overall the results show that $\pi^M$ fails to effectively balance dual objectives, overly focusing on minimizing the mobility restriction loss.

For model $\pi^S$, hospitalization rates are superior to those of $\pi^M$ in all cities except Ezhou. However, the $\bar{\mathbf{Q}}$ values in Guangzhou and Wuxi are 0.12 and 0.13, respectively, indicating that its stringent restrictions on mobility may adversely affect socio-economic activities. Similar to $\pi^M$, this strategy also fails to balance the conflicting dual objectives, as indicated by the high $D$ values across the four cities, which focus excessively on minimizing the hospital capacity strain.

\begin{table}[htbp]
  \caption{Evaluation metrics table for restriction strategies of the baselines and our method in four cities.}\label{tab:comparison_gz_wx_cq_ez}
  \centering
  \resizebox{\textwidth}{!}{%
  \begin{tabular}{cccccccc}
  \toprule
  City & Model & $\bar{H} \, \downarrow$  & $\bar{\mathbf{Q}} \, \uparrow$ & TTS $ \, \downarrow$ & ${\bar{c}} \, \downarrow$ & ${\bar{r}} \, \downarrow$ & $D \, \downarrow$ \\
  \midrule
  & No policy & 2022.27$\pm$8.21  & 1.00$\pm$0.00 & / & /&/&/ \\
  & $\pi^{T}$ & 59.05 $\pm$ 3.63 & 0.38 $\pm$ 0.04 & / & 1.82 $\pm$ 0.07&1.47$\pm$0.97&1.04$\pm$0.34\\
  Guangzhou & $\pi^M$ & 59.92 $\pm$ 1.38 & 0.59 $\pm$ 0.11 & 26.70$\pm$0.81& 1.84 $\pm$ 0.03&0.72$\pm$0.46&1.00$\pm$0.04 \\
  & $\pi^S$ & 46.79 $\pm$ 3.17 & 0.12 $\pm$ 0.03 & 27.25$\pm$1.13 & 1.53$\pm$0.06&3.03$\pm$2.11&1.18$\pm$1.80\\
  & $\pi^M_{RL}$ & 24.50 $\pm$ 1.04 &  0.54 $\pm$ 0.09 & 26.75$\pm$0.77 & 1.12$\pm$0.01&0.87$\pm$0.56&0.15$\pm$0.39\\
  & H2-MARL(ours) & $\bm{18.15\pm0.49}$ & $\bm{0.63\pm0.08}$ & $\bm{25.30\pm0.84}$& $\bm{1.03\pm0.01}$&$\bm{0.61\pm0.38}$&$\bm{0.00\pm0.00}$ \\
  \midrule
  & No policy & 776.46$\pm$3.64 &  1.00$\pm$0.00 & / & /&/&/\\
  & $\pi^{T}$ & $\bm{14.15\pm0.77}$ &  0.21 $\pm$ 0.06 & / & $\bm{0.97\pm0.01}$&2.33$\pm$1.38&0.77$\pm$1.38\\
  Wuxi & $\pi^M$ & 183.06$\pm$2.31 &  0.44 $\pm$ 0.06 & 29.9$\pm$0.62 & 10.17$\pm$0.27&1.20$\pm$0.69&1.02$\pm$0.30\\
  & $\pi^S$ & 33.21$\pm$1.04 &  0.14 $\pm$ 0.03 & 30.65$\pm$0.79 & 1.27$\pm$0.01&2.79$\pm$1.67&1.00$\pm$1.67\\
  & $\pi^M_{RL}$ & 108.68$\pm$2.78 &  0.46 $\pm$ 0.08 & 30.15$\pm$0.73 & 3.62$\pm$0.11&1.12$\pm$0.64&0.33$\pm$0.33\\
  & H2-MARL(ours) & 93.47$\pm$2.08 &  $\bm{0.56\pm0.05}$ & $\bm{26.6\pm0.92}$ & 2.93$\pm$0.07&$\bm{0.80\pm0.45}$&$\bm{0.21\pm0.07}$\\
  \midrule
  & No policy & 8440.49$\pm$7.71 &  1.00$\pm$0.00 & / & / & / & / \\
  & $\pi^{T}$ & $\bm{15.07\pm1.14}$ &  0.27 $\pm$ 0.05 & 53.95$\pm$0.80 & $\bm{0.99\pm0.01}$&1.99$\pm$0.37&0.71$\pm$0.37\\
  Chongqing & $\pi^M$ & 56.48$\pm$1.67 &  $\bm{0.60\pm0.08}$ & / & 1.75$\pm$0.03&$\bm{0.69\pm0.08}$&1.0$\pm$0.03\\
  (center district)& $\pi^S$ & 27.41$\pm$0.94 &  0.18 $\pm$ 0.03 & / & 1.17$\pm$0.02&2.53$\pm$0.52&1.02$\pm$0.51\\
  & $\pi^M_{RL}$ & 18.61$\pm$0.64 & 0.44 $\pm$ 0.07 & 21.1$\pm$0.77 & 1.04$\pm$0.01&1.20$\pm$0.18&0.28$\pm$0.17\\
  & H2-MARL(ours) & 15.97$\pm$0.61 & 0.57 $\pm$ 0.09 & $\bm{14.8\pm0.87}$ & 0.99$\pm$0.02&0.77$\pm$0.09&$\bm{0.04\pm0.09}$\\
  \midrule
  & No policy & 855.32$\pm$7.44 & 1.00$\pm$0.00 & / & / & / & / \\
  & $\pi^{T}$ & 29.67$\pm$1.11 & 0.47$\pm$0.04 & $\bm{20.05\pm0.94}$ & 1.21$\pm$0.02&1.10$\pm$0.20&0.38$\pm$0.10\\
  Ezhou & $\pi^M$ & 33.41$\pm$2.31 & 0.50$\pm$0.06 & / & 1.27$\pm$0.03&0.99$\pm$0.17&1.01$\pm$0.04\\
  & $\pi^S$ & 31.65$\pm$1.88 & 0.16$\pm$0.03 & 30.20$\pm$0.68 & 1.24$\pm$0.03&2.70$\pm$0.67&1.20$\pm$0.56\\
  & $\pi^M_{RL}$ & 29.48$\pm$1.56 & 0.47$\pm$0.07 & 22.95$\pm$0.80 & 1.20$\pm$0.02&1.10$\pm$0.20&0.29$\pm$0.13\\
  & H2-MARL(ours) & $\bm{28.02\pm1.68}$ & $\bm{0.58\pm0.05}$ & 24.00$\pm$0.62 & $\bm{1.18\pm0.02}$&$\bm{0.75\pm0.11}$&$\bm{0.00\pm0.00}$\\
  \bottomrule
  \end{tabular}}
\end{table}

The $\pi^M_{RL}$ model outperforms the first three baseline models in balancing hospitalization rates and mobility retention, evidenced by lower $D$ values across the four cities. Despite the improved performance of $\pi^M_{RL}$, it still lags behind H2-MARL in terms of $\bar{H}$, $\bar{\mathbf{Q}}$, and TTS, reflecting a lack of responsiveness and an inability to swiftly achieve more effective restriction goals. In contrast, the H2-MARL model demonstrates optimal performance, achieving successful restrictions in all four cities (no TTS = $/$) and excelling in reducing the number of hospitalizations, minimizing mobility restriction loss, rapidly reaching restriction success, and effectively balancing hospitalization capacity strain with mobility restriction loss. With the lowest metric $D$ value in all four cities, H2-MARL achieves Pareto optimality.

Overall, the H2-MARL strategy exhibits superior adaptability and effectiveness due to its balanced performance across dual objectives, making it well-suited for epidemic management in various cities.

\begin{table}[htbp]
  \caption{The simulation results of H2-MARL with different $h_0$, $l_0$.}\label{tab:generalization_experiment_H0_L0}
  \centering
  \resizebox{0.5\textwidth}{!}{%
  \begin{tabular}{ccccc}
  \toprule
  $h_0$ & $l_0$ & $\bar{H} \, \downarrow$ & $\bar{\mathbf{Q}} \, \uparrow$ & TTS $\, \downarrow$\\
  \midrule
  36&64&14.36$\pm$0.55&0.42$\pm$0.10&31.25$\pm$0.77   \\
  72&64&18.15$\pm$0.49&0.63$\pm$0.08&25.30$\pm$0.84   \\
  144&64&28.46$\pm$1.31&0.84$\pm$0.11&/   \\
  \midrule
  72&32&24.95$\pm$1.18&0.78$\pm$0.05&30.50$\pm$0.64   \\
  72&64&18.15$\pm$0.49&0.63$\pm$0.08&25.30$\pm$0.84   \\
  72&128&14.71$\pm$0.51&0.56$\pm$0.07&26.64$\pm$0.73  \\
  \bottomrule
  \end{tabular}}
\end{table}

\begin{table}[htbp]
  \caption{The simulation results of H2-MARL under epidemics with different $\beta_*^{t} (t=0)$.}\label{tab:generalization_experiment_Rt}
  \centering
  \resizebox{0.6\textwidth}{!}{%
  \begin{tabular}{ccccc}
  \toprule
  $\beta_*^{t} (t=0)$ & $R_{t,*} (t=0)$ & $\bar{H} \, \downarrow$ & $\bar{\mathbf{Q}} \, \uparrow$ & TTS $\, \downarrow$ \\
  \midrule
  0.24&1.34&7.45$\pm$0.12&0.77$\pm$0.11&17.55$\pm$0.79 \\
  0.47&2.10&18.15$\pm$0.49&0.63$\pm$0.08&25.30$\pm$0.84 \\
  0.72&2.68&29.74$\pm$1.41&0.41$\pm$0.05&28.34$\pm$0.64 \\
  \bottomrule
  \end{tabular}}
\end{table}

\begin{table}[htbp]
  \caption{The goodness-of-fit parameters $R^2$ for the simulation results of the epidemic model.}\label{tab:abaltion_D-SIHR}
  \centering
  \resizebox{0.4\textwidth}{!}{%
  \begin{tabular}{cccc}
  \toprule
  city & D-SIHR & SIHR & SIR \\
  \midrule
  Guangzhou & 0.9374 & 0.6665 &0.4201 \\
  Chongqing & 0.9546 & 0.7433 & 0.5178 \\
  Jiangsu & 0.9328 & 0.6941 & 0.5323 \\
  Hubei & 0.9787 & 0.7902 & 0.6788 \\
  \bottomrule
  \end{tabular}}
\end{table}

\subsubsection{Generalization ability}
In terms of the generalization ability of the H2-MARL, we investigate the mobility restriction strategy under varying medical conditions across mobility index and different epidemic scenarios, respectively.

Considering the differences in healthcare resources and freedom indicators among various cities, we varied the hyperparameters $l_0$ and $h_0$ with the Guangzhou dataset, while keeping other parameters consistent. $\bar{H}$, $\bar{\mathbf{Q}}$, TTS representing the generalization capability of H2-MARL are shown in Table~\ref{tab:generalization_experiment_H0_L0}, which demonstrate that H2-MARL can generate effective restriction strategies under different conditions. Specifically, a higher $h_0$ leads to higher mobility, while a higher restriction mobility penalty $l_0$ results in lower mobility. This suggests that the model's human mobility quota $\mathbf{Q}^t$ is associated with hospital capacity strain and mobility restriction loss.

Under different epidemic scenarios, we take Guangzhou as a case study to simulate different degrees of the virus by varying the parameter $\beta_*^{t} (t=0)$. The experimental evaluation results of H2-MARL are shown in Table~\ref{tab:generalization_experiment_Rt}, demonstrating that the model can generate effective restriction strategies for different scenarios. The results indicate that the model provides more lenient mobility restrictions for lower $\beta_*^t$, which represent a milder virus while enforcing stricter mobility restrictions for higher $\beta_*^t$.

\subsection{Ablation studies}
To evaluate the effectiveness of our proposed D-SIHR and H2-MARL construction, we further conduct ablation studies in this section.

\subsubsection{The ablation study for D-SIHR}
We compare the epidemic simulation accuracy of the proposed D-SIHR model with SIR and SIHR models to further address the effectiveness of D-SIHR. The ordinary differential equations (ODEs) for the SIHR model are as follows:
\begin{equation}
  \begin{aligned}
      \frac{dS}{dt} &= -\beta \frac{S I}{N}, \\
      \frac{dI}{dt} &= \beta \frac{S I}{N} - hI -  \delta I, \\
      \frac{dH}{dt} &= hI - \gamma H, \\
      \frac{dR}{dt} &= \delta I + \gamma H,
  \end{aligned}
  \label{eq: SIHR}
\end{equation}
where $\beta$, $h$, $\delta$ and $\gamma$ are defined in the same way as D-SIHR.

Using the case of COVID-19 from Section \ref{section:experiments} as an example, we set the same initial parameters (as shown in Table~\ref{tab:appendix_D-SIHR_para}). The goodness-of-fit parameters $R^2$ of the simulation results are shown in Table~\ref{tab:abaltion_D-SIHR}. The result shows that D-SIHR performs better in simulating epidemics with self-limiting characteristics. The SIR model, which does not account for mild self-recovery in self-limiting diseases, oversimplifies the description of the disease transmission process, leading to poor data-fitting results. The SIHR model fails to dynamically adjust the time-varying infection rate parameters according to the infection state and restriction strategy, resulting in significant deviations due to the constant infection rate during simulation. On the contrary, the proposed D-SIHR model shows higher accuracy by modifying the transmission mechanism and employing the online parameter update algorithm.

\subsubsection{The ablation study for H2-MARL}
For comparison, we remove the weights of the reward function in (\ref{eq: reward_function_R}), referred to as the baseline H2-NoW, and remove the expert experience replay buffer in (\ref{eq: expert_strategy_q}), referred to as the baseline H2-NoE. Parameter settings remain the same as above. We compared the evaluation metrics of epidemic human mobility restriction strategies, and the results are shown in Table~\ref{tab:abaltion_M2RL}.

All three models demonstrated effective capabilities for epidemic control in four cities. For the H2-NoW model, which lacked the adaptive adjustment mechanism for reward weights, prioritized the mobility restriction loss objective. This led to markedly increased hospital capacity strain, as evidenced by $\bar{H}$, $\bar{c}$, and TTS relative to H2-MARL while maintaining a lower mobility quota $\bar{\mathbf{Q}}$. Moreover, the H2-NoE model faced diminished sample quality and learning ability due to the lack of the expert experience replay buffer, which generates suboptimal strategies that result in a higher $\bar{H}$ and control failures in the small-scale city, i.e., Ezhou (TTS=$/$).

\begin{table}[htbp]
  \caption{Table of evaluation metrics for the restriction strategies of the model.}\label{tab:abaltion_M2RL}
  \centering
  \resizebox{\textwidth}{!}{%
  \begin{tabular}{cccccccc}
  \toprule
  City & Model & $\bar{\mathbf{Q}} \, \uparrow$ & $\bar{H}  \, \downarrow$ & TTS $ \, \downarrow$ & $\bar{c} \, \downarrow$& $\bar{r} \, \downarrow$& $D \, \downarrow$ \\
  \midrule
  & H2-MARL & 0.63$\pm$0.08 & 18.15$\pm$0.49 & $\bm{25.30\pm0.84}$ & 1.03$\pm$0.01 & 0.61$\pm$0.38 & $\bm{0.04\pm0.35}$ \\
  Guangzhou & H2-NoW & 0.31$\pm$0.08 & $\bm{14.09\pm0.56}$ & 26.25$\pm$0.77 & $\bm{0.97\pm0.01}$ & 1.81$\pm$1.21 & 1.00$\pm$1.21 \\
  & H2-NoE & $\bm{0.65\pm0.04}$ & 134.33$\pm$3.16 & 28.75$\pm$0.62 & 5.17$\pm$0.19 & $\bm{0.57\pm0.35}$ & 1.00$\pm$0.19 \\
  & No policy & 1.00$\pm$0.00 & 2022.27$\pm$8.21 & / & / & / & / \\
  \midrule
  & H2-MARL & $\bm{0.56\pm0.05}$ & $\bm{93.47\pm2.08}$ & $\bm{26.6\pm0.92}$ & $\bm{2.93\pm0.07}$ & $\bm{0.80\pm0.45}$ & $\bm{0.00\pm0.00}$ \\
  Wuxi & H2-NoW & 0.54$\pm$0.06 & 168.34$\pm$2.38 & / & 8.29$\pm$0.22 & 0.86$\pm$0.49 & 1.00$\pm$0.23 \\
  & H2-NoE & 0.41$\pm$0.04 & 109.92$\pm$3.72 & 29.70$\pm$0.71 & 3.69$\pm$0.16 & 1.32$\pm$0.76 & 1.01$\pm$0.75 \\
  & No policy & 1.00$\pm$0.00 & 776.46$\pm$3.64 & / & / & / & / \\
  \midrule
  & H2-MARL & 0.57$\pm$0.09 & $\bm{15.97\pm0.61}$ & $\bm{14.8\pm0.87}$ & $\bm{0.99\pm0.02}$ & 0.77$\pm$0.09 & $\bm{0.06\pm0.09}$ \\
  Chongqing & H2-NoW & 0.43$\pm$0.06 & 47.95$\pm$1.03 & 23.9$\pm$0.89 & 1.56$\pm$0.02 & 1.24$\pm$0.19 & 1.03$\pm$0.19 \\
  (center district)& H2-NoE & $\bm{0.58\pm0.11}$ & 106.72$\pm$2.49 & 25.50$\pm$0.74 & 3.52$\pm$0.10 & $\bm{0.74\pm0.09}$ & 1.00$\pm$0.10 \\
  & No policy & 1.00$\pm$0.00 & 8440.49$\pm$7.71 & / & / & / & / \\
  \midrule
  & H2-MARL & 0.58$\pm$0.05 & $\bm{28.02\pm1.68}$ & $\bm{24.00\pm0.62}$ & $\bm{1.18\pm0.02}$ & 0.75$\pm$0.11 & $\bm{0.03\pm0.11}$ \\
  Ezhou & H2-NoW & 0.31$\pm$0.06 & 66.78$\pm$1.04 & 30.55$\pm$0.67 & 2.02$\pm$0.03 & 1.79$\pm$0.39 & 1.01$\pm$0.39 \\
  & H2-NoE & $\bm{0.59\pm0.07}$ & 175.92$\pm$4.23 & / & 9.22$\pm$0.44 & $\bm{0.72\pm0.11}$ & 1.00$\pm$0.44 \\
  & No policy & 1.00$\pm$0.00 & 855.32$\pm$7.44 & / & / & / & / \\
  \bottomrule
  \end{tabular}}
\end{table}

In contrast, the H2-MARL model adaptively adjusted weights for the two objectives, thereby enhancing the agent's capacity to balance dual objectives effectively. The result demonstrated that the $D$ values for H2-MARL were significantly lower than those for H2-NoW and H2-NoE across all cities. Furthermore, the incorporation of expert experience in H2-MARL optimizes the learning process and avoids generating suboptimal strategies. In conclusion, the ablation study for H2-MARL highlights the critical significance of reward function weight adjustment and the expert experience replay buffer in generating the optimal strategy enhancing model performance.

\section{Conclusion}
\label{section:conclision}
In this paper, we proposed an environment simulator for infection modeling at the township level and a mobility restriction strategy model H2-MARL, achieving Pareto optimality between hospital capacity strain and human mobility during epidemic. In the environment simulator, we integrated a time-lag-free online parameter updating method into the D-SIHR infection model to enable more precise epidemic simulation. For the H2-MARL strategy model, we designed a novel reward function with dynamic adjustments of objective weights and enhanced the exploration process through expert experience replay buffer and heuristic spatial pruning. We further established a large-scale OD dataset to support the mobility restriction research. Finally, our comprehensive experiments validate the effectiveness, scalability, and adaptability of the proposed D-SIHR model and H2-MARL strategy model. The results demonstrate the superior performance of H2-MARL in dual-objective decision-making, confirming its potential for practical application in urban epidemic scenarios.



The research on urban epidemic simulation and control presented in this paper has significant practical implications for the development of smart health cities. It serves as an essential component of building a comprehensive disaster simulation system for smart cities. Finally, it should be noted that the specific implementation methods of mobility restriction strategy for each AD are beyond the scope of this study, and we recommend considering this as a topic for subsequent research. This may involve the development of specific mobility management plans for urban roads, public places, and factories.

\section*{Acknowledgements}
This work was supported in part by the National Key Research and Development Project under Grant 2019YFB2102300, in part by the National Natural Science Foundation of China under Grant U23A20382, and in part by the Natural Science Foundation of Shanghai under Grant 23ZR1467300.


\bibliographystyle{elsarticle-harv}

\bibliography{lxt-refs}




\end{document}